\documentclass[]{aastex631}
\usepackage{natbib}
\usepackage{amsmath}
\usepackage{verbatim}
\usepackage{amssymb}
\usepackage{hyperref}
\usepackage{float}
\usepackage{graphicx}
\usepackage{mathtools}
\usepackage{empheq}
\usepackage{appendix}
\usepackage{microtype}
\usepackage{commath}
\usepackage{bm}
\usepackage{bold-extra}

\DeclareGraphicsExtensions{.pdf,.png,.jpg}

\def\arcsec{\ensuremath{^{\prime\prime}}}

\newcommand{\kmsmpc}{\hbox{$ \, \rm km\, s^{-1} \, Mpc^{-1}$}}

\newcommand{\bq}{\begin{equation}} 
\newcommand{\eq}{\end{equation}} 
 
\newcommand{\beq}{\begin{equation}}
\newcommand{\eeq}{\end{equation}}
\newcommand{\beqa}{\begin{eqnarray}}
\newcommand{\eeqa}{\end{eqnarray}}

\newcommand{\PL}{$P$--$L$\ }
\newcommand{\PLs}{$P$--$L$\ }

\shorttitle{Crowded No More}
\shortauthors{Riess et al.}

\begin{document} 

\title{Crowded No More: The Accuracy of the Hubble Constant Tested\\with High Resolution Observations of Cepheids by {\it JWST}} 

\author[0000-0002-6124-1196]{Adam G.~Riess}
\affiliation{Space Telescope Science Institute, 3700 San Martin Drive, Baltimore, MD 21218, USA}
\affiliation{Department of Physics and Astronomy, Johns Hopkins University, Baltimore, MD 21218, USA}

\author[0000-0002-5259-2314]{Gagandeep S. Anand}
\affiliation{Space Telescope Science Institute, 3700 San Martin Drive, Baltimore, MD 21218, USA}

\author[0000-0001-9420-6525]{Wenlong Yuan}
\affiliation{Department of Physics and Astronomy, Johns Hopkins University, Baltimore, MD 21218, USA}

\author{Stefano Casertano}
\affiliation{Space Telescope Science Institute, 3700 San Martin Drive, Baltimore, MD 21218, USA}

\author{Andrew Dolphin}
\affiliation{Raytheon, 1151 E. Hermans Road, Tucson, AZ 85706}

\author[0000-0002-1775-4859]{Lucas M.~Macri}
\affiliation{NSF's NOIRLab, 950 N Cherry Ave, Tucson, AZ 85719, USA}

\author[0000-0003-3889-7709]{Louise Breuval}
\affiliation{Department of Physics and Astronomy, Johns Hopkins University, Baltimore, MD 21218, USA}

\author[0000-0002-4934-5849]{Dan Scolnic}
\affiliation{Department of Physics, Duke University, Durham, NC 27708, USA}

\author{Marshall Perrin}
\affiliation{Space Telescope Science Institute, 3700 San Martin Drive, Baltimore, MD 21218, USA}

\author[0000-0001-8089-4419]{Richard I.~Anderson}
\affiliation{Institute of Physics, Laboratory of Astrophysics, \'Ecole Polytechnique F\'ed\'erale de Lausanne (EPFL),\\ Observatoire de Sauverny, 1290 Versoix, Switzerland}

\begin{abstract}

High-resolution {\it JWST} observations can test confusion-limited {\it HST} observations for a photometric bias that could affect extragalactic Cepheids and the determination of the Hubble constant. We present {\it JWST} NIRCAM observations in two epochs and three filters of $>$320 Cepheids in NGC$\,$4258 (which has a 1.5\% maser-based geometric distance) and in NGC$\,$5584 (host of SN~Ia 2007af), near the median distance of the SH0ES {\it HST} SN Ia host sample and with the best leverage among them to detect such a bias. {\it JWST} provides far superior source separation from line-of-sight companions than {\it HST} in the NIR to largely negate confusion or crowding noise at these wavelengths, where extinction is minimal. The result is a remarkable $>$2.5$\times$ reduction in the dispersion of the Cepheid \PL relations, from 0.45 to 0.17~mag, improving individual Cepheid precision from 20\% to 7\%. Two-epoch photometry confirmed identifications, tested {\it JWST} photometric stability, and constrained Cepheid phases. The \PL relation intercepts are in very good agreement, with differences {\it (JWST$-$HST)} of 0.00$\pm$0.03 and 0.02$\pm$0.03~mag for NGC$\,$4258 and NGC$\,$5584, respectively. The difference in the determination of H$_0$ between {\it HST} and {\it JWST} from these intercepts is 0.02$\pm$0.04~mag, insensitive to {\it JWST} zeropoints or count-rate non-linearity thanks to error cancellation between rungs. We explore a broad range of analysis variants (including passband combinations, phase corrections, measured detector offsets, and crowding levels) indicating robust baseline results. These observations provide the strongest evidence yet that systematic errors in {\it HST} Cepheid photometry do not play a significant role in the present Hubble Tension. Upcoming {\it JWST} observations of $>$12 SN~Ia hosts should further refine the local measurement of the Hubble constant.
\end{abstract}

\section{Introduction} \label{sec:intro}

\begin{figure}[b]
\begin{center}
\includegraphics[width=0.8\textwidth]{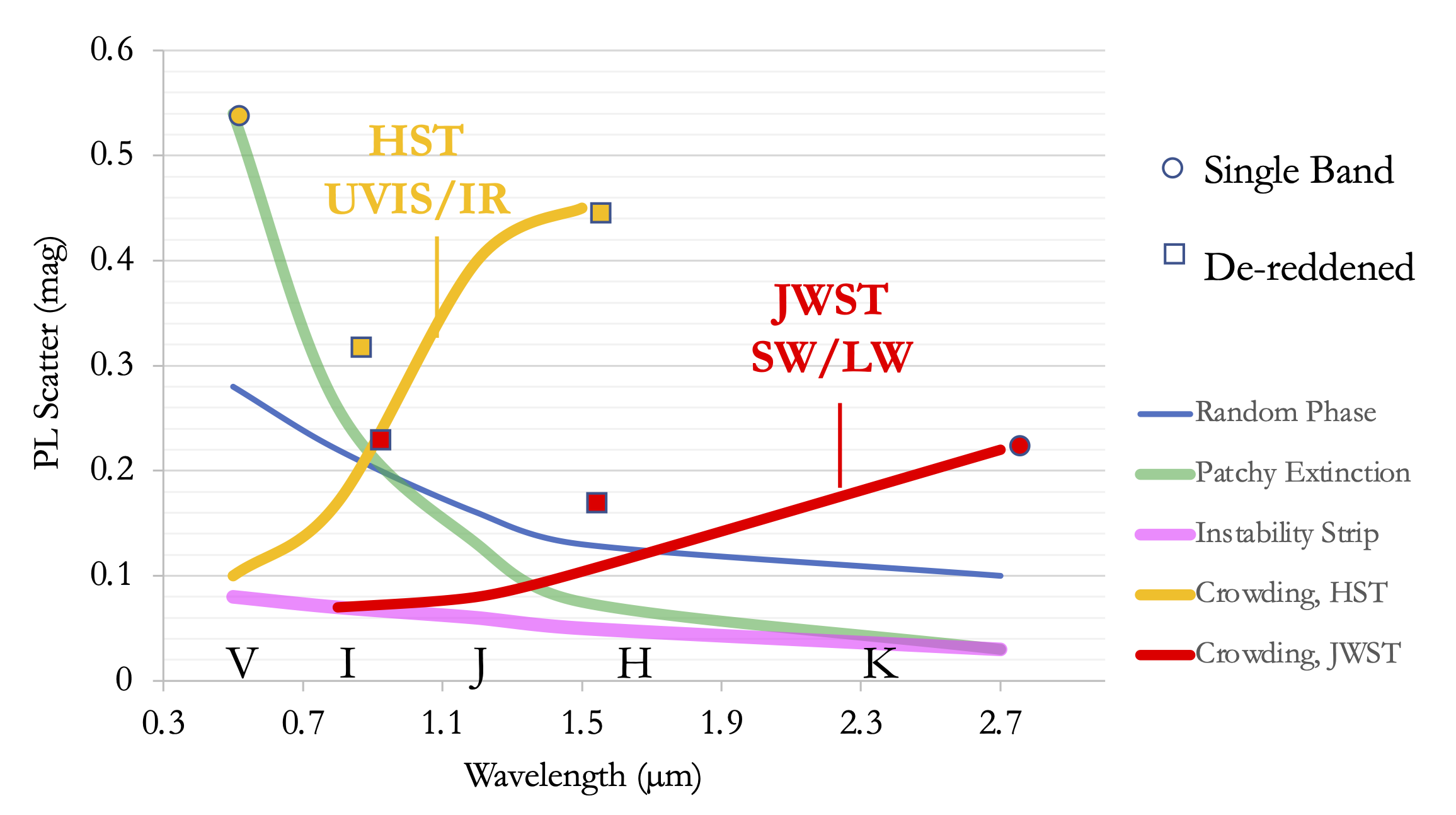}
\end{center}
\caption{\label{fg:scatter} Estimated sources of scatter in extragalactic Period-Luminosity relations as indicated (add in quadrature).  Scatter from random phase sampling and patchy extinction can be reduced through multi-epoch sampling and color measurements and are dominant in the optical.  Crowding is an a important source of scatter in the NIR, but is greatly reduced by the resolution of {\it JWST}.  These sources do not include photon statistics and assume low galaxy inclination.  Measurements from {\it HST} and {\it JWST} for NGC$\,$5584 and 4258 are indicated as gold and red symbols, respectively.}
\end{figure}

Over the last decade an intriguing tension has emerged between the value of the Hubble constant (H$_0$) measured directly from redshifts and distances, following its local definition and independent of models, and the same parameter inferred from the cosmological model, calibrated in the early Universe (\citealp[for recent reviews]{Verde:2019,Divalentino:2021,Kamionkowski:2022}).  This ``Hubble Tension'' is seen with high significance ($>$5$\sigma$ confidence) across a wide-range of local, independent distance indicators and among experiments measuring the Cosmic Microwave Background (CMB) and thus requires a solution. The {\it James Webb Space Telescope (JWST)} provides new capabilities to scrutinize and refine some of the strongest observational evidence for this tension.

The most significant differences are seen between measurements of local Type Ia supernovae (SNe~Ia) calibrated by Cepheid variables, which yield $H_0=73.0 \pm 1.0$ \kmsmpc\ \citep[][hereafter R22]{Riess:2022}, and the analysis of {\it Planck} observations of the Cosmic Microwave Background \citep{Planck2018}, which predict $H_0=67.4 \pm 0.5$ \kmsmpc\ in conjunction with $\Lambda$CDM. Cepheids have been the preferred primary distance indicator for these studies because they are extraordinarily luminous ($M_H\sim-$7~mag at a period of 30 days), intrinsically precise ($\sim$3\% in distance per star), reliably identifiable due to their periodicity \citep[since][]{Leavitt:1912}, and well understood \citep[since][] {Eddington:1917}.  They are also the best-calibrated distance indicator accessible in a large volume of SN~Ia hosts thanks to consistent use of a single, stable instrument, {\it HST} WFC3 UVIS+IR, for all measurements in SN hosts and in several independent geometric anchors: the megamaser host NGC$\,$4258, the Milky Way (through parallaxes, now including Gaia EDR3), and the LMC and SMC (via detached eclipsing binaries). Near-infrared (NIR) observations are essential to mitigate the impact of dust, the bane of many cosmic probes.   

However, all known long-range distance indicators have deficiencies. For measurements of Cepheids in SN~Ia hosts at D$\geq$10 Mpc, it is often not possible, even with the resolution of  {\it HST}, to fully separate these variables from their stellar crowds.  As \citet{Freedman:2019} noted, ``[p]ossibly the most significant challenge for Cepheid measurements beyond 20 Mpc is crowding and blending from redder (RGB and AGB) disk stars, particularly for near-infrared H-band measurements [...]”.  The background levels of individuals Cepheids result from the randomness of source superposition, the inclination of the host galaxy, and our line of sight. The solution to the problem of crowding (also called blending or confusion) has been to trade precision for accuracy by quantifying the {\it mean} level of the local background due to brightness fluctuations using ``artificial star'' photometry.  Artificial stars of pre-defined brightness may be randomly added to the images near Cepheids and their recovered magnitudes used to account for the mean background of real stars.   Numerous tests show (see \S4 for a discussion) the resulting Cepheid magnitudes are accurate but at the cost of greatly reduced individual precision of each Cepheid, as shown in Fig.~\ref{fg:scatter} which details the sources of Cepheid \PL uncertainties.  Fortunately, the mean precision of a host's Cepheid sample is still greater than a single-hosted SN~Ia provided a sample of $>$25 Cepheids per host, a condition exceeded by 75\% of the SH0ES SN~Ia hosts.

The successful launch of {\it JWST} \citep{Gardner2023, Rigby2023} provides the means to separate Cepheids from their stellar crowds, thanks to its unparalleled resolution, and recover their intrinsic precision.  Cepheids are generally identified and their periods measured with {\it HST} at optical wavelengths, where the amplitudes of their light curves are 3-4$\times$ greater than in the NIR, facilitating their discovery.  Follow-up observations at longer wavelengths mitigate dust effects, but with higher background and lower resolution.  In the NIR, observations with {\it JWST} reduce the contaminating background by a factor of $\sim$8 with respect to {\it HST}, a result of the increase in the square of the ratio of the effective resolutions at 1.5$\mu$m (also abetted by WFC3-IR's undersampled pixels) and the accompanying decrease in the area blended with the Cepheid PSF.  The resolution of {\it JWST} in the NIR is comparable to that of {\it HST} at visible wavelengths, providing {\it JWST} the advantage of both low extinction and high resolution in the same images. Our Cycle 1 {\it JWST} program (GO 1685) is observing the 6 richest Cepheid hosts ($\sim$100 variables per host) of 9 SN~Ia in the SH0ES sample (R22) and the maser host NGC$\,$4258.  Here we present the first replication of the {\it HST} Cepheid-SN~Ia distance ladder using high resolution observations with {\it JWST}. We present the observations in \S2, our analysis in \S3 and the discussion in \S4.
\begin{deluxetable*}{ccccccccc}[t]
\tabletypesize{\scriptsize}
\tablecaption{Observation Log\label{tb:obs}}
\tablewidth{0pt}
\tablehead{
\colhead{Date} & \colhead{Epoch} & \colhead{Exposure$^a$} & \colhead{Filter1} & \colhead{Filter2} & \colhead{Exp. time [s]} & \colhead{RA (J2000)} & \colhead{Dec (J2000)} & \colhead{Orientation}
}
\startdata
2023-01-30 &    N5584e1 &            009001\_03101\_* &  F090W &  F277W &    418.7$\times4$ &    215.59944 &     -0.38796 &    287.6 \\
2023-01-30 &    N5584e1 &            009001\_05101\_* &  F150W &  F277W &    526.1$\times4$ &    215.59944 &     -0.38796 &    287.6 \\
2023-02-21 &    N5584e2 &            010001\_02101\_* &  F090W &  F277W &    418.7$\times4$ &    215.59948 &     -0.38793 &    282.6 \\
2023-02-21 &    N5584e2 &            010001\_02103\_* &  F150W &  F277W &    526.1$\times4$ &    215.59948 &     -0.38793 &    282.6 \\
2023-05-02 &    N4258e1 &            005001\_03101\_* &  F090W &  F277W &    257.7$\times4$ &    184.75304 &     47.37235 &    140.2 \\
2023-05-02 &    N4258e1 &            005001\_03103\_* &  F150W &  F277W &    365.1$\times4$ &    184.75304 &     47.37235 &    140.2 \\
2023-05-17 &    N4258e2 &            006001\_03101\_* &  F090W &  F277W &    257.7$\times4$ &    184.70268 &     47.33871 &    136.1 \\
2023-05-17 &    N4258e2 &            006001\_03103\_* &  F150W &  F277W &    365.1$\times4$ &    184.70268 &     47.33871 &    136.1
\enddata
\tablecomments{$a$: All exposures start with jw01685.}
\end{deluxetable*}

\section{Observations}
The goal of  the reported {\it JWST} observations is to measure a large sample of Cepheids using positions and periods measured by {\it HST} in SN~Ia hosts and the maser host, NGC$\,$4258, to complete a distance ladder.   As with the SH0ES program (R22), this ladder by design negates photometric zeropoint uncertainties through cancellation between the maser and SN~Ia hosts.  

\begin{figure}[b]
\begin{center}
\includegraphics[width=0.8\textwidth]{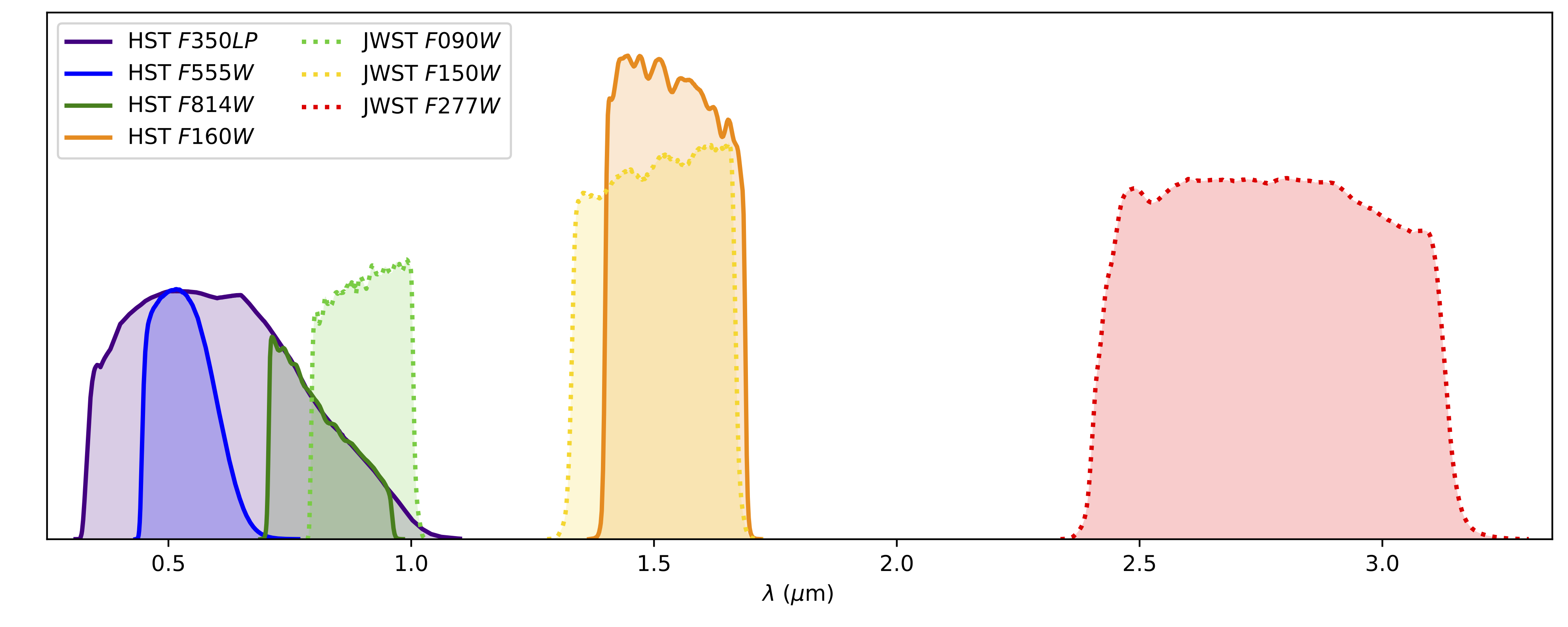}
\end{center}
\caption{\label{fg:filters} {\it HST} and {\it JWST} Bandpass systems referenced in this work.  The {\it HST} filters are $F555W,F814W,F350LP,$ and $F160W$, and the {\it JWST} filters include $F090W,F150W,$ and $F277W$.  There is a good correspondence between $F814W$ and $F090W$ and a very good correspondence between $F160W$ and $F150W$.}
\end{figure}

\begin{figure}
\centering
\plottwo{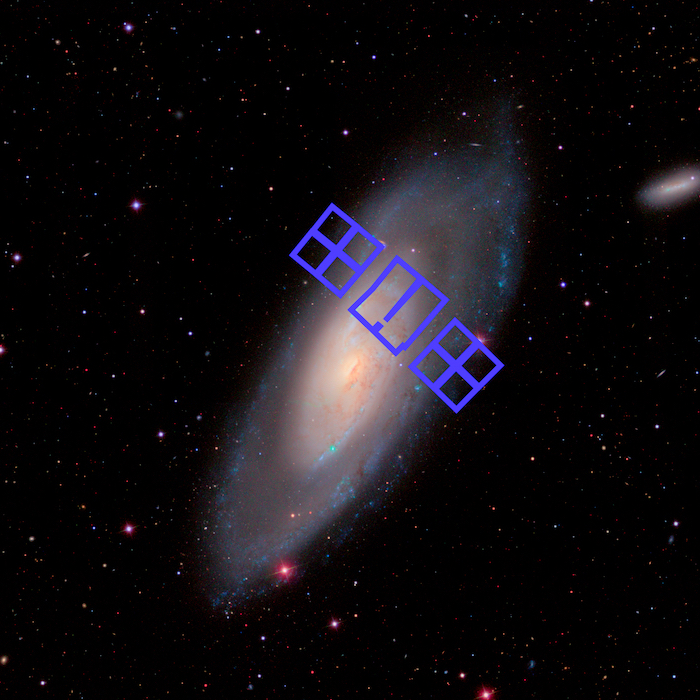}{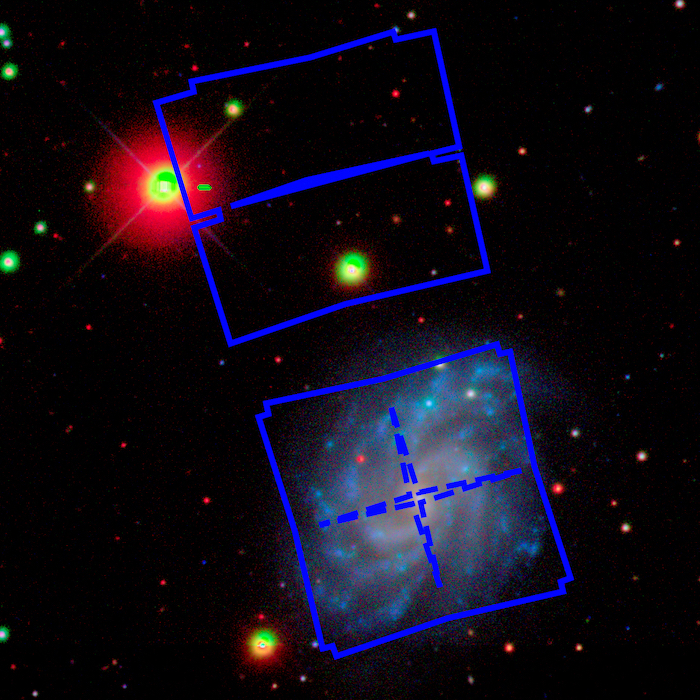}
\caption{\label{fg:n4258_image} NIRCAM fields superimposed on SDSS \textit{gri} images \citep{SDSS2015} of NGC 4258 (left) and NGC 5584 (right). North is up and east is to the left.}
\end{figure}

\begin{figure}
\centering
\plottwo{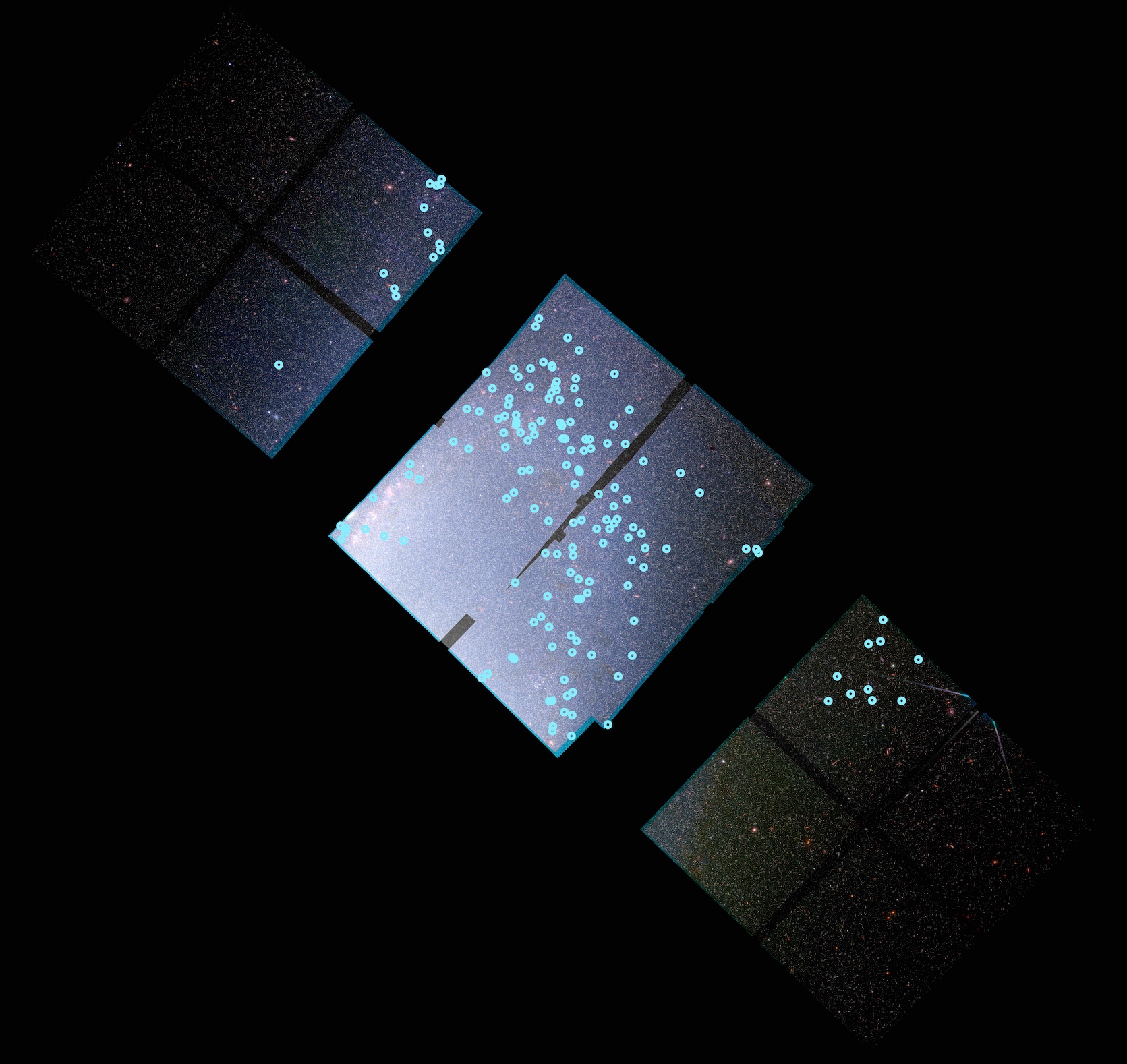}{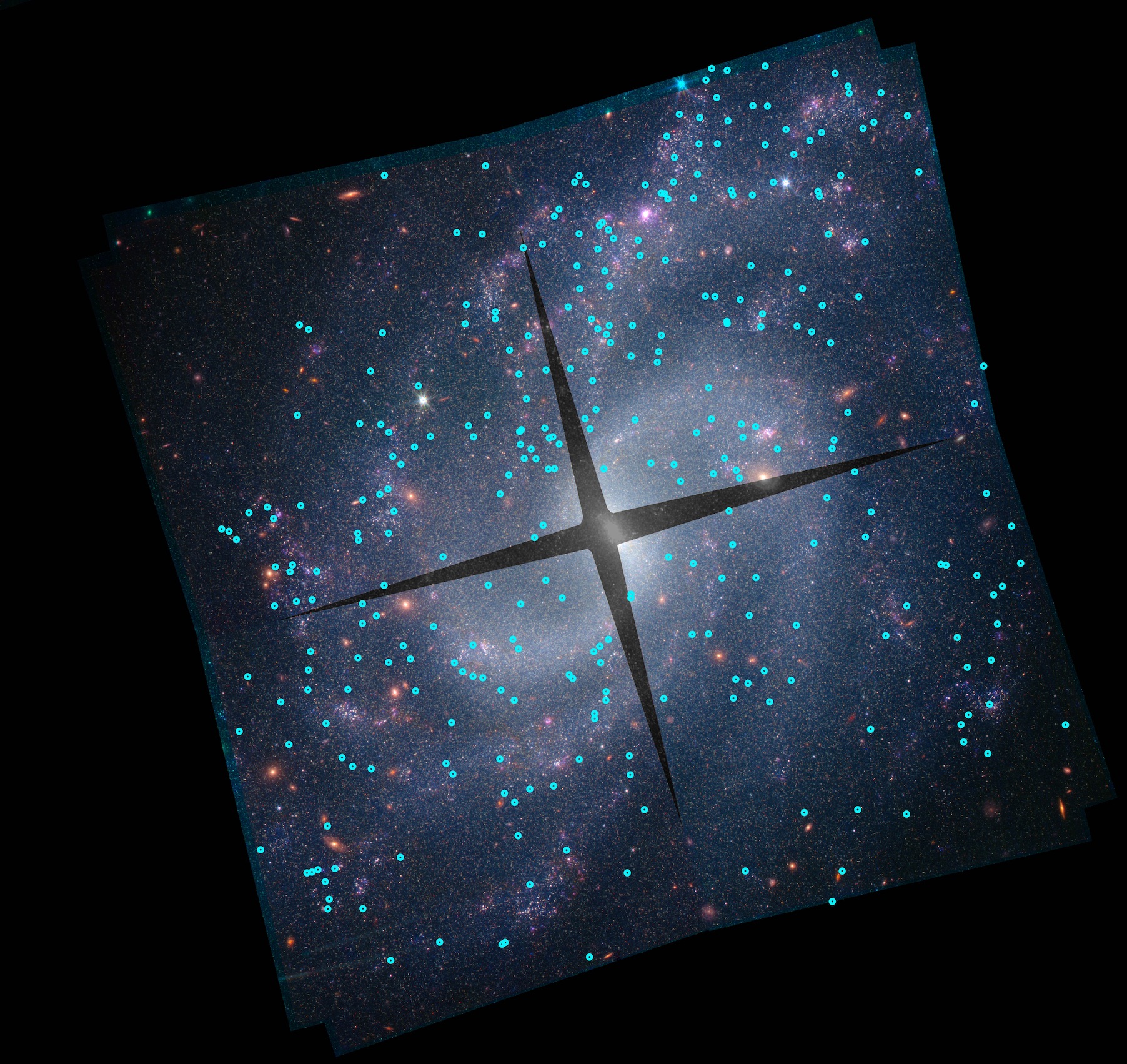}
\caption{\label{fg:n5584_image} NIRCAM RGB images (F090W/F150W/F277W) showing positions of Cepheids (cyan circles). Regions without short-wavelength coverage (due to the module gaps) were manually set to grayscale.}
\end{figure}

The observations were taken during the first half of 2023 with {\it JWST} NIRCAM modules A and B for each host at two epochs and in three filters, $F090W$ (0.9$\mu$m), $F150W$ (1.5$\mu$m), and $F277W$ (2.8$\mu$m) as described in Table 1. The filters used to observe the Cepheids with {\it HST} and {\it JWST} are compared in Fig.~\ref{fg:filters}.  The regions imaged for the two hosts are shown in Figures \ref{fg:n4258_image} and \ref{fg:n5584_image}. The filter $F150W$ was chosen because it is extremely close to {\it HST} $F160W$, facilitating a very direct comparison between the two space platforms.  $F090W$ largely overlaps {\it HST} $F814W$. $F277W$ is unexplored territory for Cepheids, and indeed much of space-based astronomy,  but offers low extinction, minimal PAH background, and some (useful) sensitivity to individual Cepheid metallicity through the presence of CO bandheads \citep{Scowcroft:2016}.  Two epochs of observations were chosen to: (a) statistically confirm the astrometric identification of the Cepheids by their variability, (b) recover phase information, and (c) check the photometric stability of {\it JWST} NIRCAM over time spans relevant to this program. 

\begin{figure}[b]
\begin{center}
\includegraphics[width=0.99\textwidth]{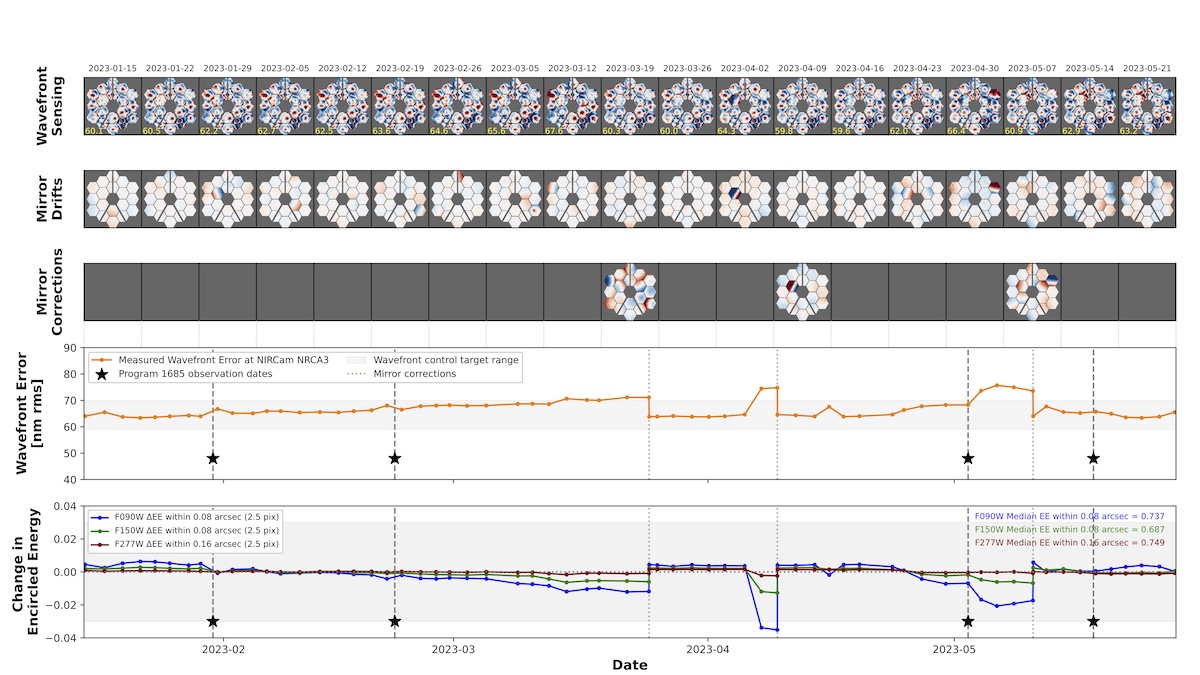}
\end{center}
\caption{\label{fg:wavefront} {\it JWST} wavefront sensing data during the interval of these observations in the first half of 2023.  The wavefront sensing, drifts, remediations and wavefront errors are shown in the top four panels.  The bottom panel shows the change to encircled energy in the core of the PSF in the three passbands used for this program.  The dates of our observations are indicated as vertical dashed lines and a star symbol.}
\end{figure}

We verified that no significant telescope wavefront changes occurred during the course of these observations that would affect point source photometry at $>1\%$. This analysis made use of measurements from the Cycle 1 telescope wavefront sensing monitoring program \citep{Lajoie2023} available via the MAST archive. We retrieved the telescope monitoring data for the period of these observations, 2023 January through May, including all measured drifts and commanded mirror corrections. The telescope was fairly stable during this time period, with a few corrections applied during April and May to maintain alignment better than 70 nanometers (see \citealt{Lajoie2023} for further discussion of JWST's stability). From the time series wavefront data we computed model PSFs using WebbPSF \citep{Perrin:2014}, measured PSF encircled energies, and assessed the variations in encircled energy over that time period (see Fig.~\ref{fg:wavefront}). Based on this analysis, systematic effects on PSF photometry due to telescope alignment variations are limited to $<$1\% for F090W, $<$0.25\% for F150W, and $<$0.12\% for F277W for the observing dates in this program. Even these changes are reduced by the use of frame-specific aperture corrections as described below.

We use the STScI NIRCAM reduction pipeline data, version 1.10.1, and reference {\it JWST} file version 1084.pmap to ensure consistency of the frame calibration across the program observations. 

\subsection{DOLPHOT Photometry}

We perform PSF photometry using the NIRCAM module\footnote{We use the April 6, 2023 version of the module and the PSFs, which are the latest available at the time of this work.} of the DOLPHOT software package \citep{Dolphin:2000, DOLPHOT2016}. While the module itself is still currently in a ``beta" version, it has been developed and tested based on the {\it JWST} Early Release Science program \citep{Weisz:2023, Weiszinprep}. Additionally, the underlying algorithms used within DOLPHOT have been well tested by many science programs over the last two decades \citep{Karachentsev2002, Dalcanton2009, Radburn-Smith2011, Williams2014, McQuinn2017, Anand2021}.  We have utilized several recent versions of the pipeline and DOLPHOT and have found these to change the mean Cepheid magnitude of a large sample at the $\sim$0.01 magnitude level. 

While we anticipate updates to the calibration and characterization of {\it JWST}, these are unlikely to impact the results presented here because the quantities we present are based on {\it differences} in Cepheid photometry between two hosts which greatly reduce, if not fully negate, these sources of photometric uncertainty.  However, we recognize several known issues which may impact {\it absolute} photometry at the few percent level, such as count rate non-linearity (likely to be very close to ideal but not yet measured), chip-specific zeropoints and their temporal variation, improvements to knowledge of the PSF and more subtle effects that can impact photometry. Based on monitoring data, the NIRCAM Team at STScI has seen a downward trend of about $\sim$1\% in sensitivity over the prior year, which might indicate our observations are brighter at a fraction of that level (M. Boyer, private communication). Based on recent iterations, we estimate a systematic uncertainty in our {\it absolute} photometry of $\sigma$=0.03 mag. However, absolute uncertainties are less relevant for the photometric comparison of Cepheids across distance ladder rungs which depends only on {\it relative} photometry.

We perform the photometry on the stage 2 \textit{*cal.fits} images, using a combined stage 3 \textit{*i2d.fits} as the reference frame for mutual image alignment and deep source detection. These images are obtained directly from the Mikulski Archive for Space Telescopes (MAST). We tested using both the $F090W$ and $F150W$ filters as reference frames. In these tests, we found slightly better object detection with $F150W$ (likely due to its improved throughput), and chose to use it as the reference frame going forward. We also tested the effects of a de-striping procedure on the underlying images (using the methods of \citealt{Yuan2022JWST}). We found a negligible difference on the final intercept of the Cepheid PLR ($<$0.005 mag, including testing of both reference frames), and thus chose to perform photometry on the images obtained directly from MAST to simplify future comparisons. 

Additionally, in our initial tests, we found that performing photometry with both the SW ($F090W$ and $F150W$) and LW channels ($F277W$) based on a simultaneous source list produced from all wavelengths resulted in SW photometry that was systematically offset from cases where SW photometry was measured based on a source list constructed only from SW data. Upon examination of the source-subtracted images, 
it became clear that the source detection in the combined SW+LW data was inferior to the use of SW-only data, likely due to the poorer resolution of the LW images which appears to ``veto'' sources resolved at SW. Poorer source detection in the combined SW+LW photometry resulted in biased aperture corrections (due to incomplete removal of sources from the background), and thus biased photometry. To avoid this issue, we performed all of the SW photometry on only the SW ($F090W$ and $F150W$) images without reference to the LW image. Once the SW measurements are complete, we use the source list from this SW photometry for the combined SW+LW photometry using the ``warmstart" option within DOLPHOT. After inspection of the residuals from this warmstart run, we find that the residuals of the PSF photometry are much closer to the ideal case when performing SW photometry alone. However, there are uncommon instances where some point-source flux is still not appropriately subtracted in the SW data. Hence, in our final results, we use the $F090W$ and $F150W$ photometry from the SW-only run, and the $F277W$ photometry from the combined SW+LW run which uses the best option for each. 

When running DOLPHOT, we adopt the parameters recommended in the NIRCAM module's manual\footnote{\url{http://americano.dolphinsim.com/dolphot/dolphotNIRCAM.pdf}}. A significant portion of the initial photometry list from DOLPHOT contains detections of a low quality.  For our photometric quality cuts, we use a modified version of the cuts provided in \cite{Warfield2023}. Specifically, we select for sources which meet the following criteria in the SW data: (1) Crowding $<$ 0.5; (2) $\mathrm{Sharpness^{2}}$ $\le$ 0.01; (3) Object Type $\le$ 2; (4) S/N $\ge$ 3; (5) Error Flag $\le$ 2. All of the cuts are applied on both $F090W$ and $F150W$ filters (except for object type, which is not a filter-specific output). We also use the value of the $\chi^2$ statistic of the source fits as reported by DOLPHOT as an additional quality cut as described in the next section.  
\subsection{Cepheid Astrometric Matching}

Cepheid catalogs were previously produced from the SH0ES team multi-epoch, optical {\it HST} images  \citep{Riess:2022,Yuan:2022a}. We derived linear transformations between {\it HST} and {\it JWST} frames using all the sources with small photometric errors ($\sigma <$0.1~mag), then applied these transformations to Cepheids and derived their positions in the {\it JWST} frames.  The position error predicted from this method is marginal compared to the position errors from individual {\it HST} Cepheid centroid measurements and is generally $\sim$0.2 pixels.

\begin{figure}[t]
\plottwo{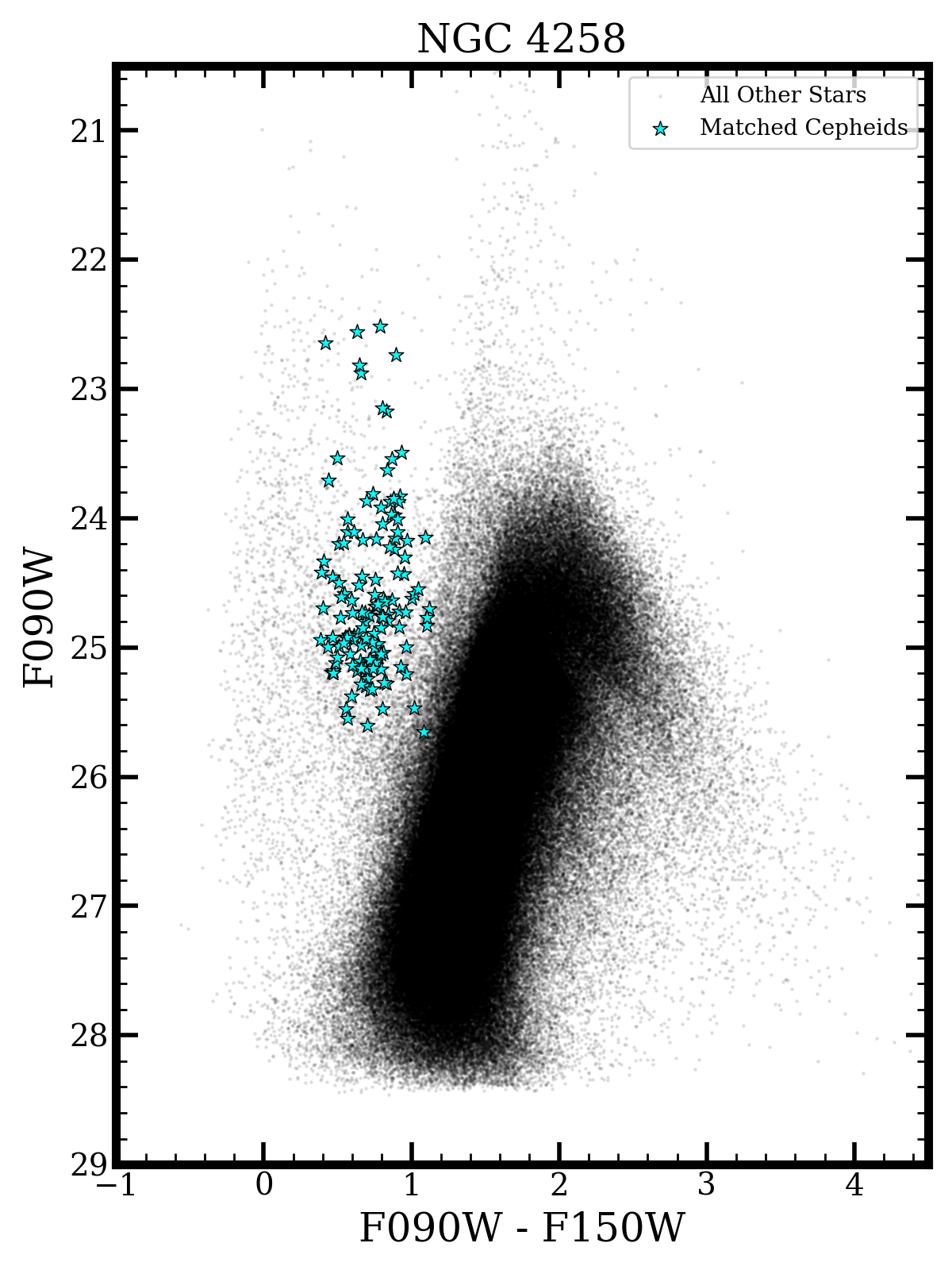}{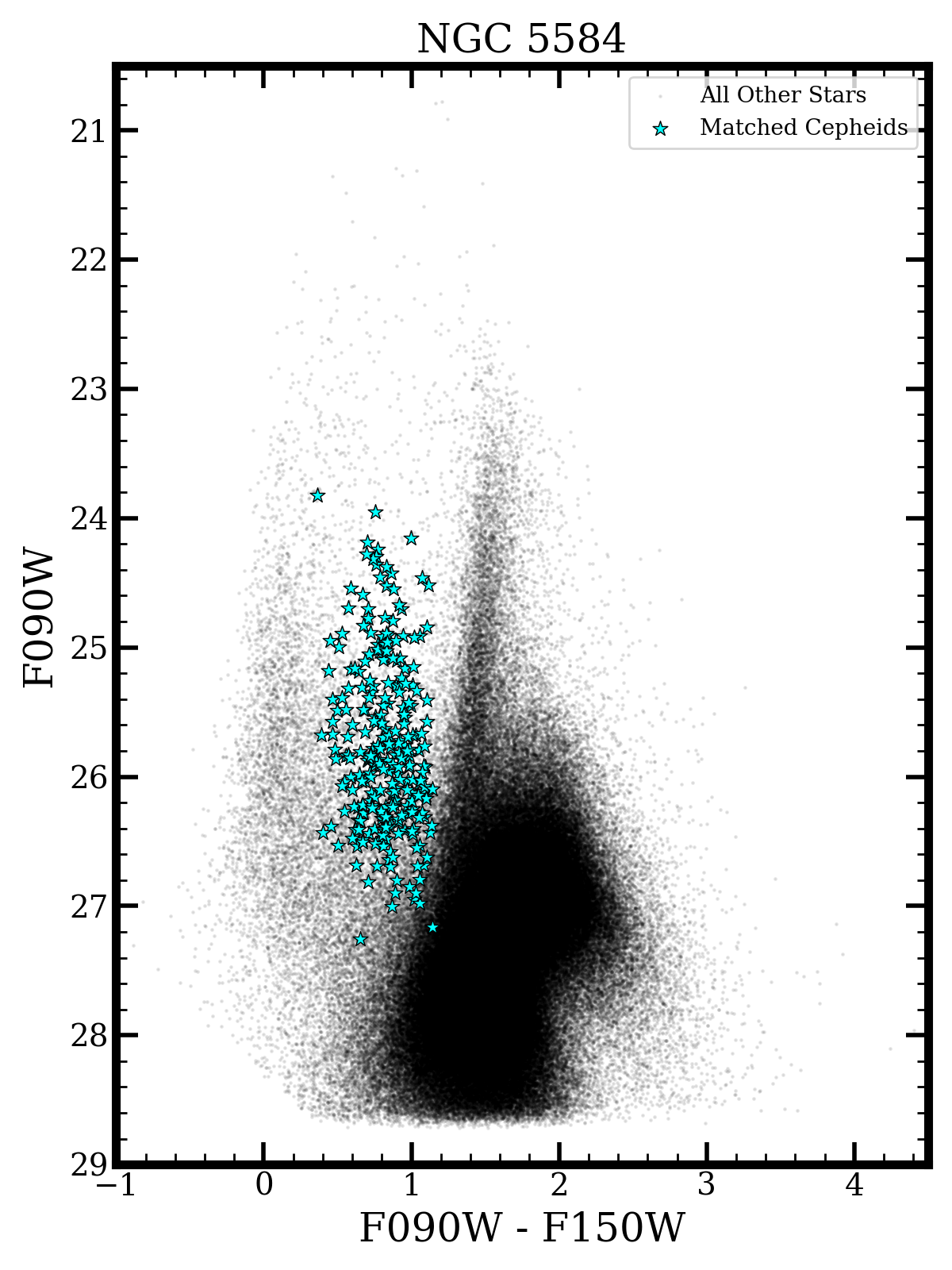}
\caption{\label{fg:cmds} Color magnitude diagrams for the modules covering Cepheids.  Each star has a small black point while Cepheids are indicated by a cyan star.  A color cut in this space is included, $0.3 < F090W-F150W < 1.15$ corresponding to a broad range around the instability strip ($0.5 < V-I < 1.7$). The background CMDs are shown from the first visit. If a Cepheid is observed in only the first or both visits, its location on the CMD is shown from the first visit. Otherwise, if it is observed in only the second visit, its location is shown from the second visit.}
\end{figure}

\begin{figure}[ht]
\begin{center}
\includegraphics[width=0.85\textwidth]{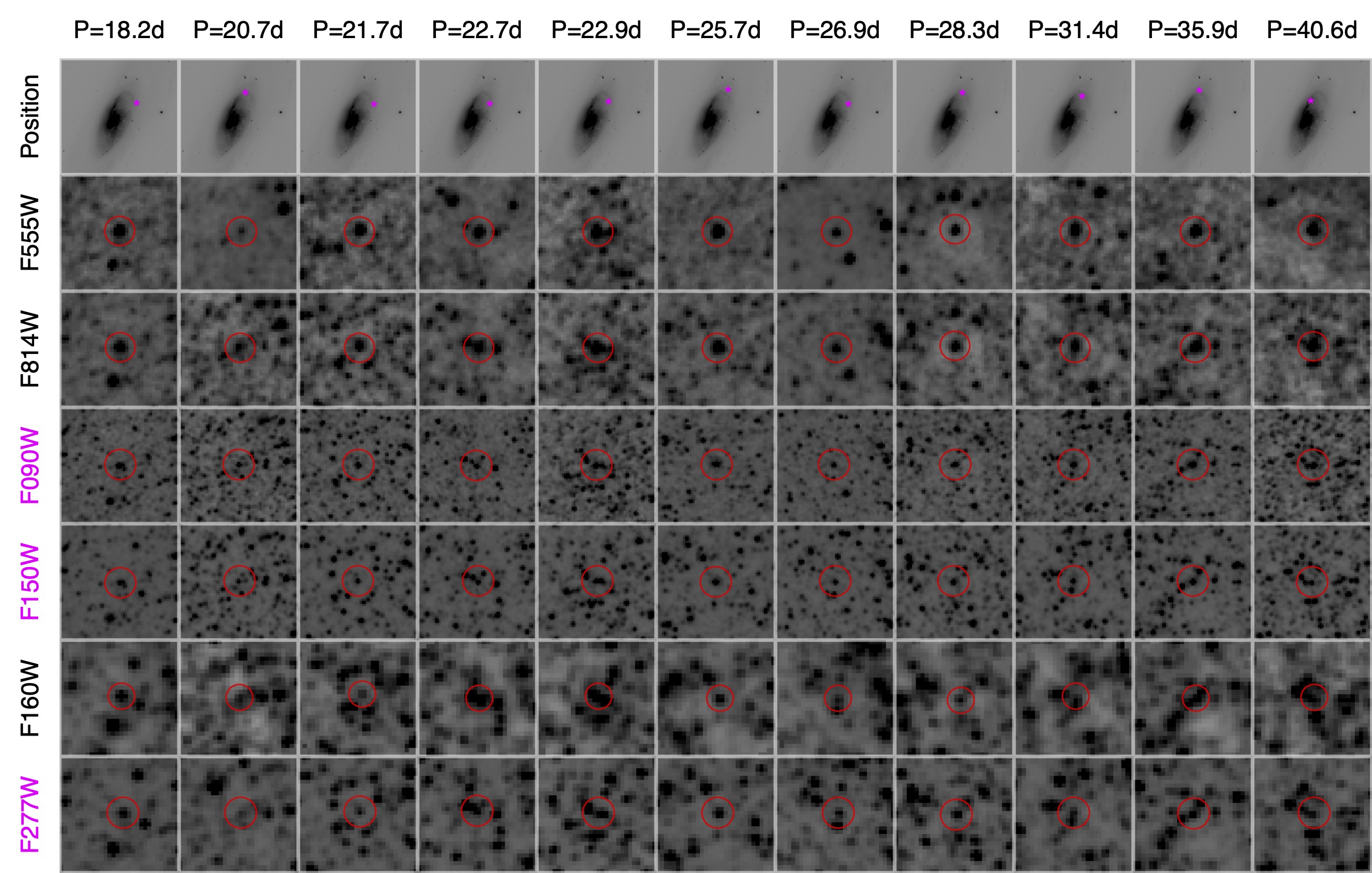}
\end{center}
\caption{\label{fg:stamps4258} {\it HST} and {\it JWST} image stamps in NGC$\,$4258 for all Cepheids with period of 18 to 41 days. The upper row indicates the location of each Cepheid.  {\it HST} filters are labeled in black and {\it JWST} filters in magenta.}
\end{figure}

\begin{figure}[ht]
\begin{center}
\includegraphics[width=0.85\textwidth]{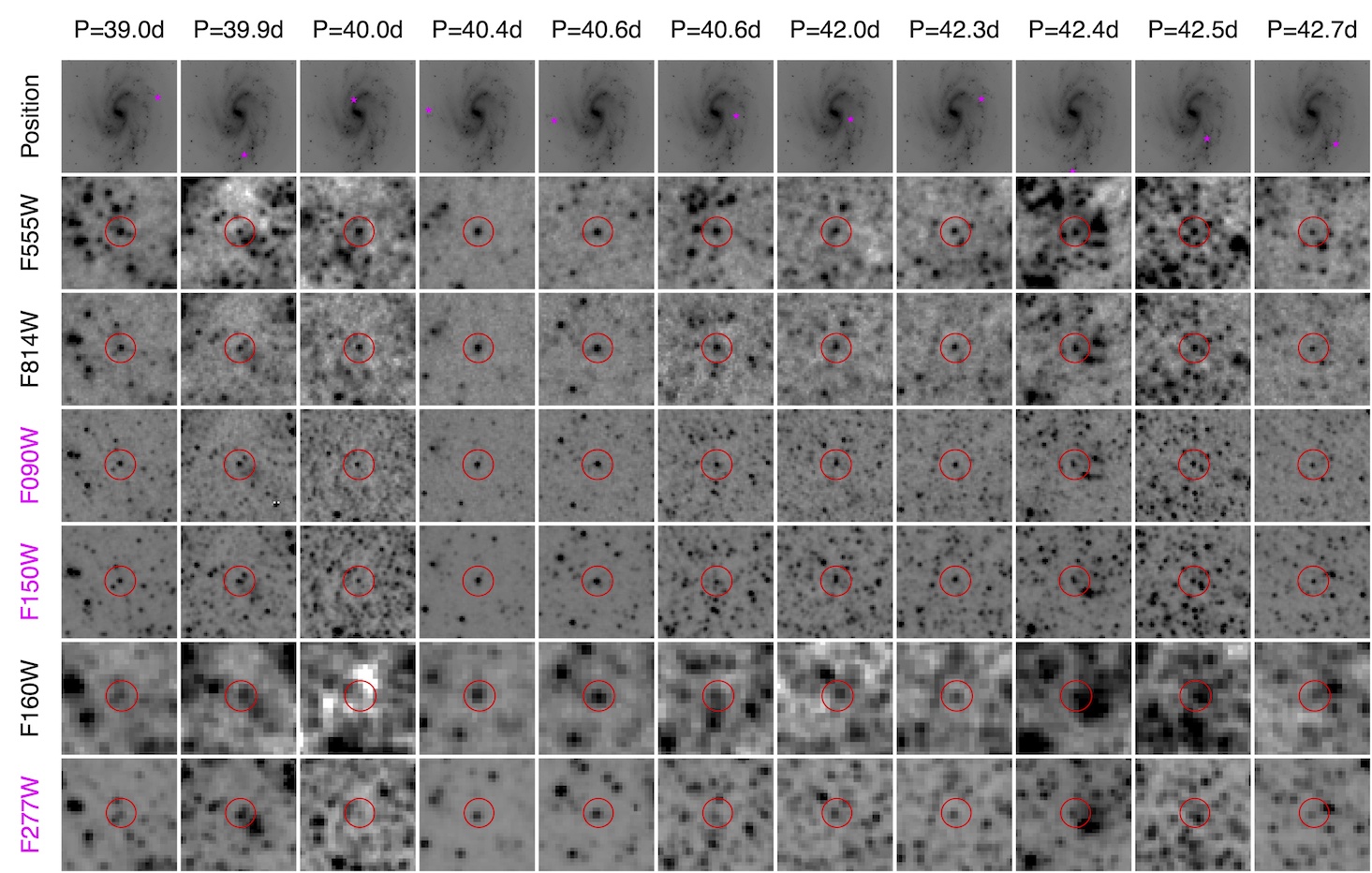}
\end{center}
\caption{\label{fg:stamps5584} Same as Fig.~\ref{fg:stamps4258} but for NGC$\,$5584, and with periods of 39 to 43 days.}
\end{figure}

With high-quality {\it JWST} photometry and predicted Cepheid positions determined, we identify the Cepheids within the {\it JWST} data. We use a k-d tree to find the nearest match for each predicted Cepheid in the culled DOLPHOT source list. We only consider the matching successful if there is a source found within 0.02\arcsec or 0.707~SW pixels ($0.5^{2}$, a tolerance of 0.5 pixels in each axis) of the predicted position.  In practice, matches were obtained with a median matching distance of $\sim$0.2 pixels, consistent with the individual uncertainties in the {\it HST} centroids. We further verify the accuracy of our procedure by examining the positions of the matched Cepheids on the color-magnitude diagrams (CMDs) of the host galaxies, as the matched sources will lie in the instability strip of the CMD.  The recovered Cepheids are shown with the underlying CMD is in Fig.~\ref{fg:cmds} in the expected regions.  Small image cutouts (stamps) for Cepheids in the {\it HST} ($F555W$, $F814W$, $F160W$) and {\it JWST} bandpasses are shown in Figures \ref{fg:stamps4258} and \ref{fg:stamps5584}. We provide our matched Cepheid photometry in Table \ref{tb:phot}.

 To ensure reliable Cepheid magnitudes, we require a good value of $\chi^2$ (i.e., the modeling of the local scene) as provided by DOLPHOT, $<$1.4 per degree of freedom for the mean Cepheid.  A high value generally indicates a bad fit to the PSF or superposition of PSFs and we find $\sim$5-10\% of sources fail this cut across these hosts \footnote{Even for well-isolated sources, $\chi^2$ may be high for brighter sources because of imperfections in the PSF model which become more apparent and hence significant with decreasing shot noise.  For this reason we made this quality cut a modest linear function of log period (which corresponds to Cepheid brightness but avoids the bias of referencing the apparent brightness), varying from 1.3 for short periods to 1.6 for long periods, as determined empirically from the location of \PL outliers.}. We use a color cut of $0.3 < F090W-F150W < 1.15$, equivalent to $0.5 < V-I < 1.7$, as stars outside this range are either strongly reddened, blended with a red or blue star, or misidentified, with about 5\% of sources failing this cut.  With future, larger {\it JWST} samples it may be possible to refine these color ranges in the {\it JWST} system.

\subsection{Artificial Stars}

While our {\it JWST} NIR imaging has about three times the resolution at the same wavelength as {\it HST}, the level of crowding present near the Cepheids is still not completely negligible nor can be ignored. To quantify the effects of crowding on the {\it JWST} Cepheid photometry, we perform injection and recovery simulations, i.e. artificial star tests within DOLPHOT. For each matched Cepheid and in each filter, we insert and attempt to recover 100 stars of the same magnitude as the measured Cepheid, as well as 100 stars with an input magnitude 0.3 magnitudes fainter than the measured Cepheid. We can interpolate between these two injection values for the crowding-corrected Cepheids. To quantify the local environment of each Cepheid, we place each one of these artificial Cepheids in a random location within 1 arcsecond of the measured position, with a 5 pixel exclusion radius right around the Cepheid (the exclusion radius is essential to avoid a bias due to the unequal condition that fake stars could be superimposed by chance on the known, nearby Cepheid, which will not occur for the actual Cepheid). Additionally, we do not put down an artificial star if the randomly determined position falls directly onto a bad pixel or off the detector (as such Cepheids would not be included in our sample due to the quality cuts), instead trying to place that artificial star down again in a re-generated position. Notably, these artificial stars are placed and recovered one at a time, as to not increase the intrinsic level of crowding seen within the images.  

For NGC$\,$5584, the mean background bias (i.e., crowding correction) derived from the artificial stars is 0.03, 0.07 and 0.08~mag and the random errors due to crowding (derived from the dispersions of the individual sets of artificial stars) are 0.08, 0.13, 0.21~mag in F090W, F150W, and F277W, respectively.  The values are very similar for NGC$\,$4258, 0.03, 0.06 and 0.08 for the background bias and 0.07, 0.13, 0.20 for the random errors.  The similarity of these values means the bias has negligible impact on the distance determination for the SN~Ia host; forgoing crowding corrections altogether for both hosts increases the net distance measured to the SN~Ia host by 0.008 mag.

It is important to apply the same quality cuts, color cuts and inclusion criteria to the artificial stars used to derive the mean background (or crowding) level or else this mean, applied to the Cepheids, will be biased.  We verify for both the optically-identified Cepheids and the artificial stars, 80-85\% are retained, with most that are lost due to the same reason; superposition on a complex region resulting in DOLPHOT reporting a poor fit or lack of recovery\footnote{These rejected Cepheids, apparent with {\it HST} in the optical due to the good contrast at short wavelengths, are superimposed on dense concentrations of red giants which are not well resolved even with JWST.}.

\startlongtable
\begin{deluxetable*}{cccccccccccccc}
\tabletypesize{\scriptsize}
\tablewidth{0pt}
\tablenum{2}
\tablecaption{Photometric Data for Cepheids\label{tb:phot}}
\tablehead{\colhead{Host} &  \colhead{ID} & \colhead{ra} & \colhead{dec}  & \colhead{Log P} & \colhead{$F090W$} & \colhead{$\sigma$}  & \colhead{$F150W$} & \colhead{$\sigma$} & \colhead{$F277W$} & \colhead{$\sigma$} & \colhead{$V-I^a$} & \colhead{$\sigma$} & \colhead{note}} 
\startdata
N5584 &        96196 &  215.58141  &  -0.38763  &  1.2434  &  26.175  &  0.115  &  25.549  &  0.111  &  25.367  &  0.140  &  0.957  &  0.117  & note \\
N5584 &       114600 &  215.58182  &  -0.39029  &  1.2317  &  26.505  &  0.130  &  25.765  &  0.129  &  25.641  &  0.159  &  0.897  &  0.158  & note \\
N5584 &       115209 &  215.58295  &  -0.38981  &  1.2015  &  26.598  &  0.138  &  25.940  &  0.173  &  25.783  &  0.198  &  0.817  &  0.157  & note \\
N5584 &       134727 &  215.58605  &  -0.39115  &  1.2538  &  26.364  &  0.159  &  25.480  &  0.186  &  25.245  &  0.207  &  1.126  &  0.189  & note \\
\hline
\enddata
\tablecomments{$a$: $F555W-F814W$.}
\phantom{}
\vspace{-48pt}
\end{deluxetable*}
\section{Analysis}
\subsection{Epoch differences}

A simple mean of magnitude differences (epoch 1 - epoch 2) demonstrates the stability of {\it JWST} across epochs for (NGC$\,$5584, NGC$\,$4258);

   \begin{itemize}
    \item $F090W$=(-0.011$\pm$0.018, -0.001$\pm$0.023) mag
    \item  $F150W$=(-0.009$\pm$0.011, 0.018$\pm$0.013) mag
    \item $F277W$=(-0.004$\pm$0.014, -0.031$\pm$0.015) mag
   \end{itemize}
   
\begin{figure}[b]
\begin{center}
\includegraphics[width=0.68\textwidth]{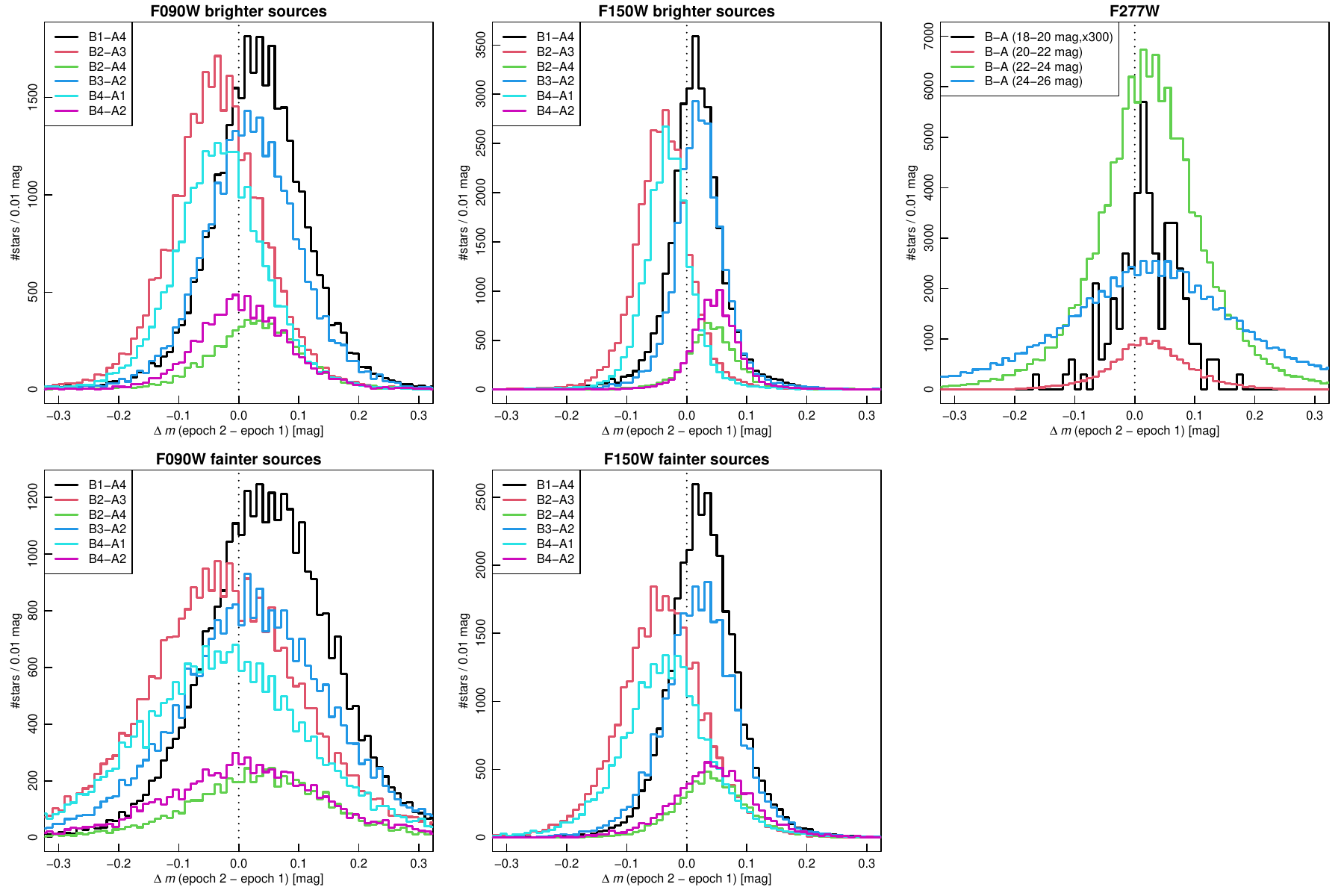}
\end{center}
\caption{\label{fg:chiptochip} Differences in the photometry of stars observed in two epochs on different positions of the NIRCAM field.}
\end{figure}

In the case of NGC$\,$5584, the Cepheids in each epoch generally land on the same chip and module.  For the case of NGC$\,$4258 they land on different chips and modules between the two epochs. In the case of  NGC$\,$4258, we offset the pointing between the two epochs such that the same Cepheid-rich disc field was observed first with Module A and later by Module B.  This shift provided the opportunity to compare the non-variable stars and measure chip-vs-chip and module-vs-module differences.  It also ensures we can compare Cepheid magnitudes between the anchor NGC$\,$4258, and SN~Ia hosts with the same NIRCAM module no matter which is used for the latter.  We show the chip versus chip and module vs module differences derived from the two epochs for non-variable stars in Fig.~\ref{fg:chiptochip}.

These chip-to-chip differences for the SW vary from 0.01 to 0.05~mag with a standard deviation of 0.03 mag\footnote{Implied monochromatic offsets to minimize the variance across all chips in magnitudes evaluated right to left: A1=A1-0.03; B4=B4; A2=A2+0.008; B3=B3-0.016;  A3=A3-0.046; B2=B2+0.023; A4=A4+0.055; B1=B1+0.012 mag}.  We are  uncertain whether these are true differences in chip zeropoints,  making them applicable across all observations, or instead reflect uncertainties in aperture corrections and thus would be specific to these observations.  We do not apply these offsets in our baseline analysis. However, in the next section we consider changes to the distance estimates if we make use of these (which we find to be small, at the $\sim$0.01~mag level). For $F277W$ the comparison of the two modules (epoch 1-epoch 2) gives a median difference of -0.026 mag, similar to the -0.031~mag for the Cepheids.

\subsection{Cepheid Phasing}

\begin{figure}[t]
\plottwo{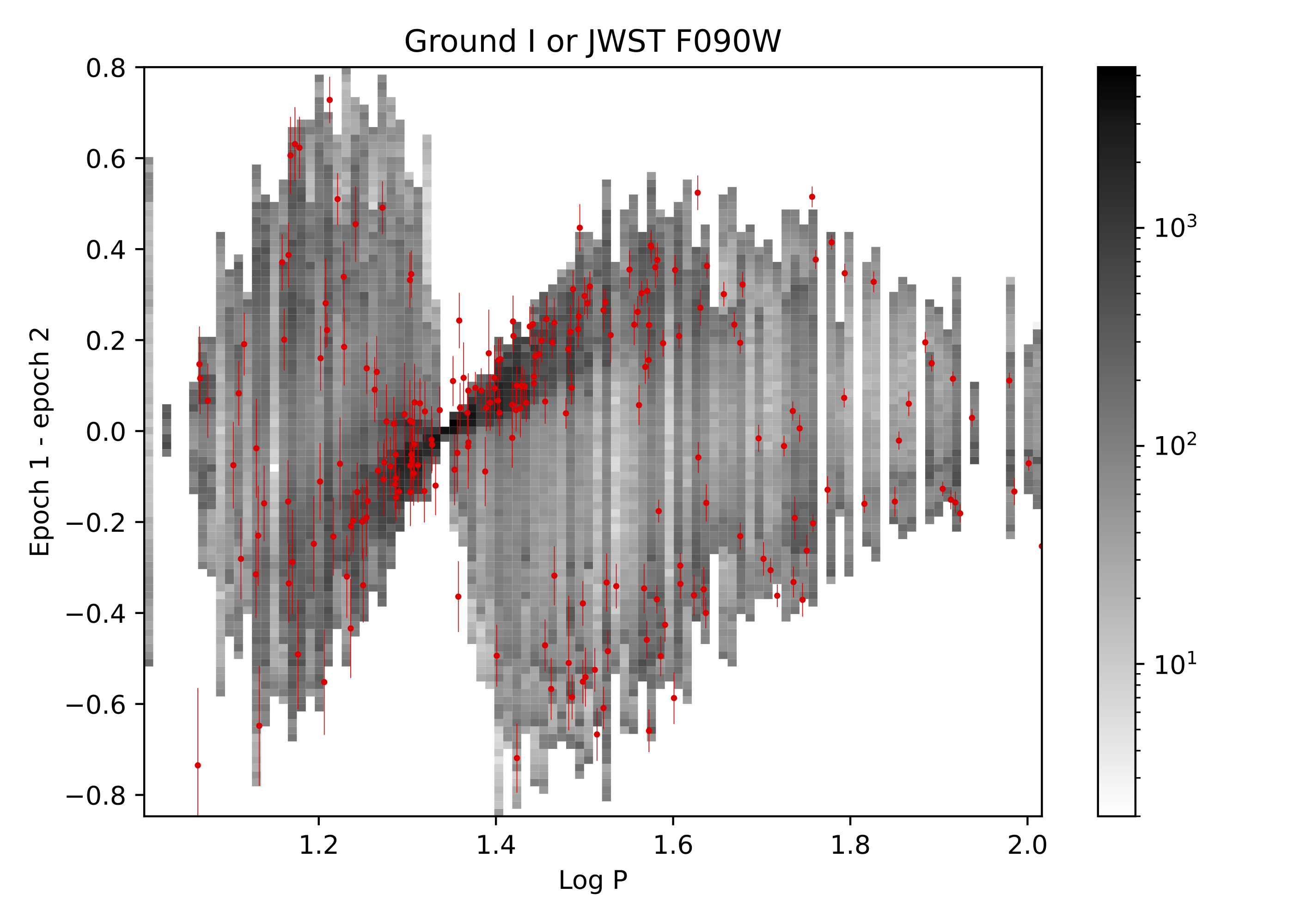}{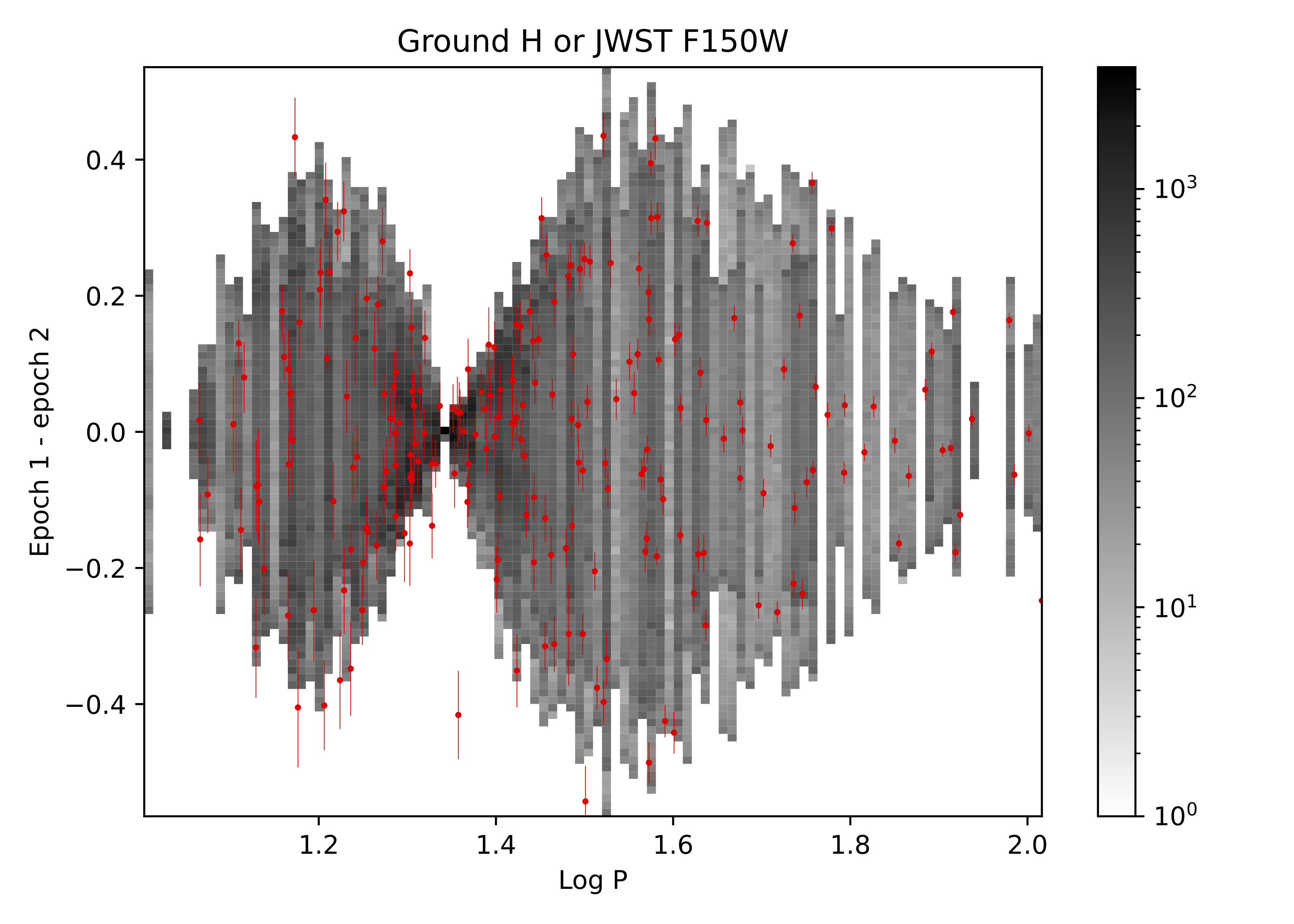}
\caption{\label{fg:deltamag5584} Magnitude differences of two epochs for Cepheids in NGC$\,$5584,  Left is $F090W$ and right is $F150W$.  Gray density shows expected frequency of sampling based on random phase and template light curves.  Red points are measurements from {\it JWST} and the errors do not include the crowding error which effectively cancels in the difference.  The asymmetry of Cepheid light curves in $F090W$ produces structure in this diagram that can be used to constrain the phase. The log of the time interval between epochs produces a negligible difference at a value of $\log P=$1.34.}
\end{figure}

\begin{figure}[h]
\plottwo{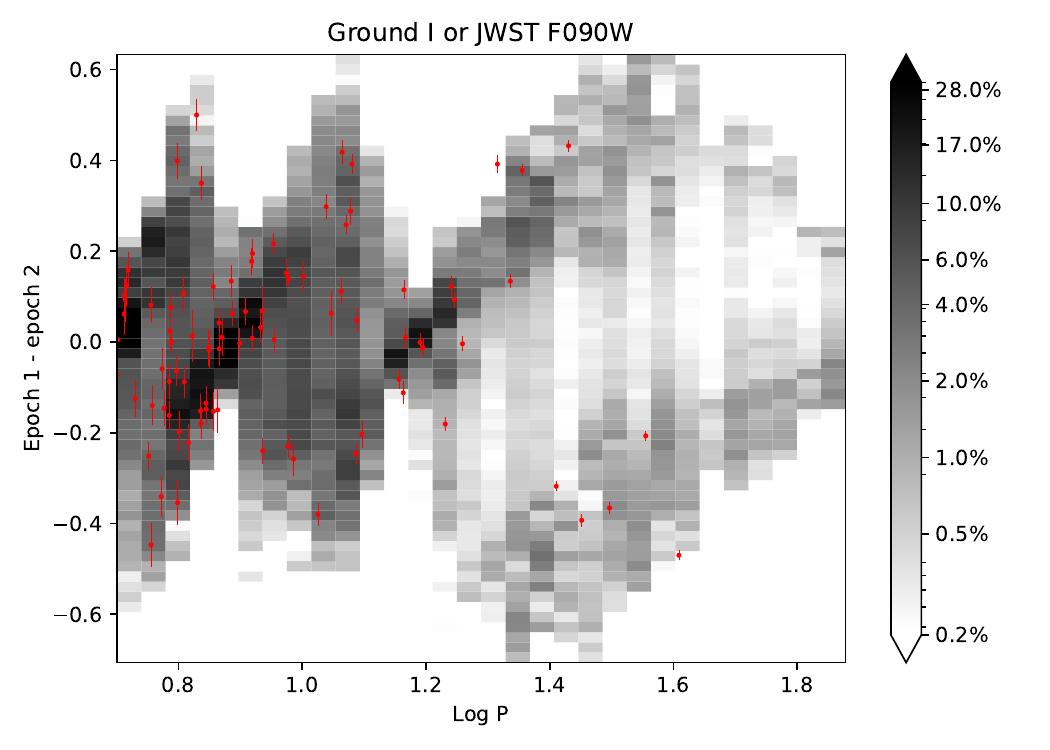}{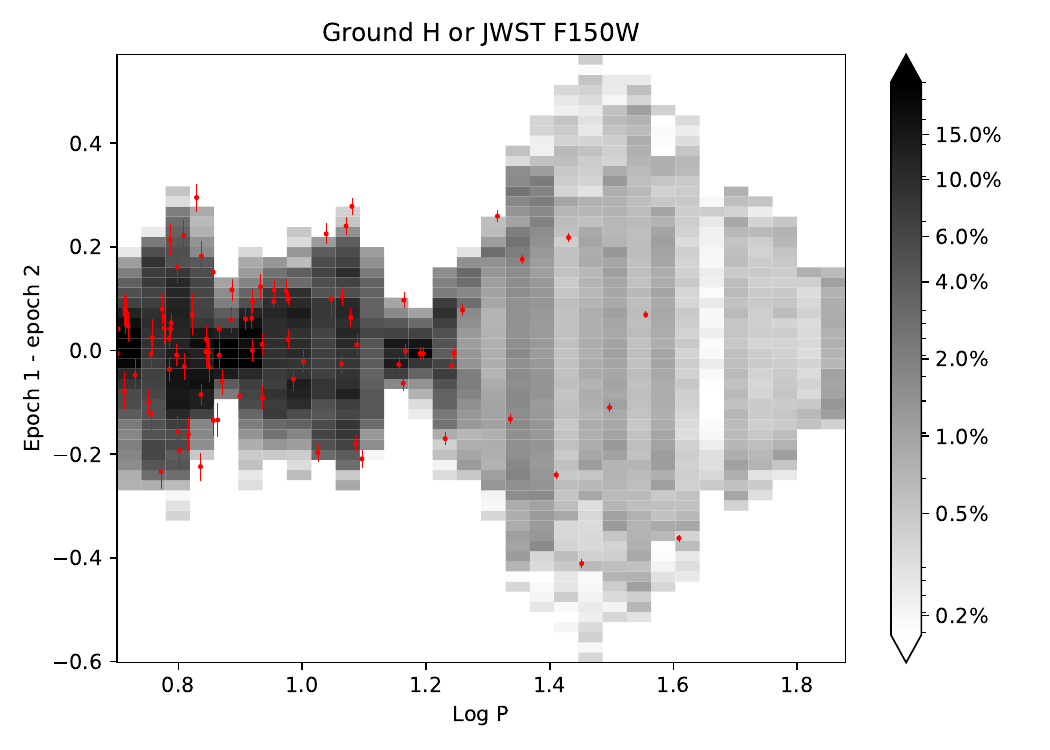}
\caption{\label{fg:deltamag4258} Same as Fig.~\ref{fg:deltamag5584} but for NGC$\,$4258, with an epoch time difference of $\log P=$1.18.}
\end{figure}

The precision of the {\it JWST} Cepheid photometry is apparent from the epoch differences shown versus their periods in Figures \ref{fg:deltamag5584} and \ref{fg:deltamag4258} and in comparison to the expectations from random sampling of template Cepheid light curves.  The symmetry of the $F150W$ difference diagram is a consequence of the symmetric light curve (around the mean) of Cepheids in the NIR.  The ``saw-toothed'' light curves of Cepheids in the optical results in the asymmetric profile seen in $F090W$.  

At the fixed time separation between the pairs of epochs (22.14 days for NGC$\,$5584 and 15.27 days for NGC$\,$4258), the Cepheid period corresponds directly to a calculable difference in phase of the two light curve samples. The Cepheid phase of the first epoch may be considered {\it random} because any constraint on the phase has elapsed since their {\it HST}-based discovery.  Specifically, the period determinations have a mean precision of $\sim$2\% \citep{Yuan:2021}, so after a few years the phase error is of order unity.

For periods where the time separation is a multiple of the period we expect a magnitude difference of near zero as the two samples are at the same phase.  For periods halfway between a multiple,  a maximum difference, twice the amplitude of the light curve, occurs for some Cepheids, depending on their specific phases.  The simulated light curves (gray in Figures \ref{fg:deltamag5584} and \ref{fg:deltamag4258}) are in excellent agreement with the measurements and indicate the ability to constrain the phase for periods where the phase difference is large.   

\begin{figure}[b]
\epsscale{1.0}
\plottwo{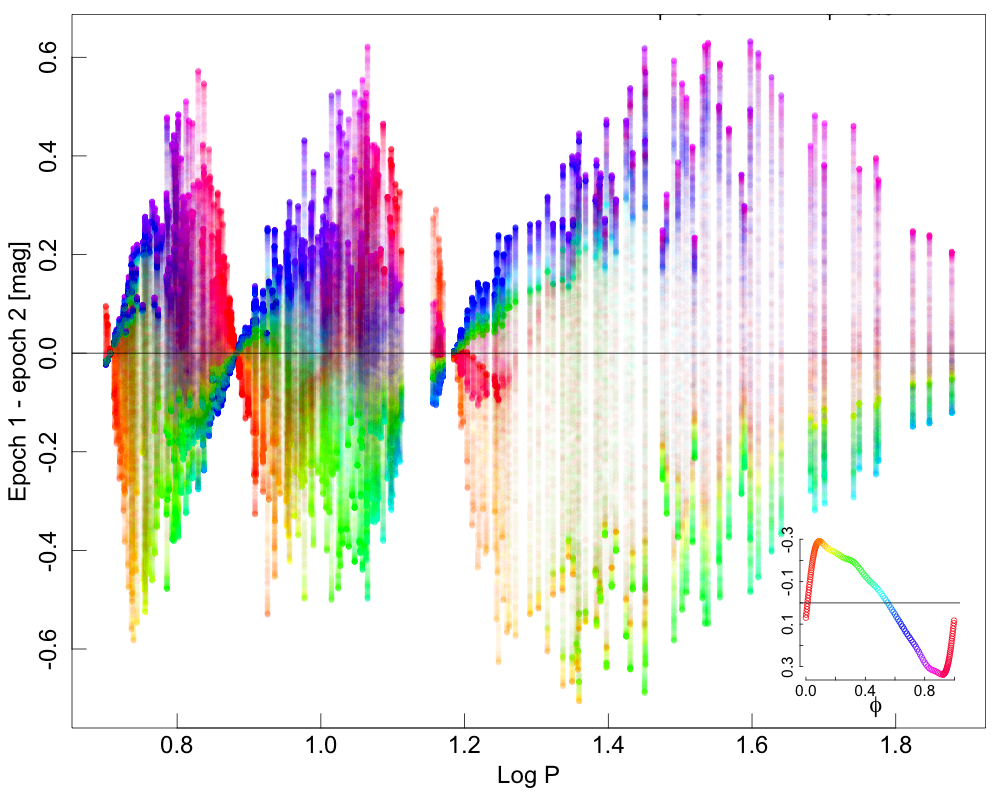}{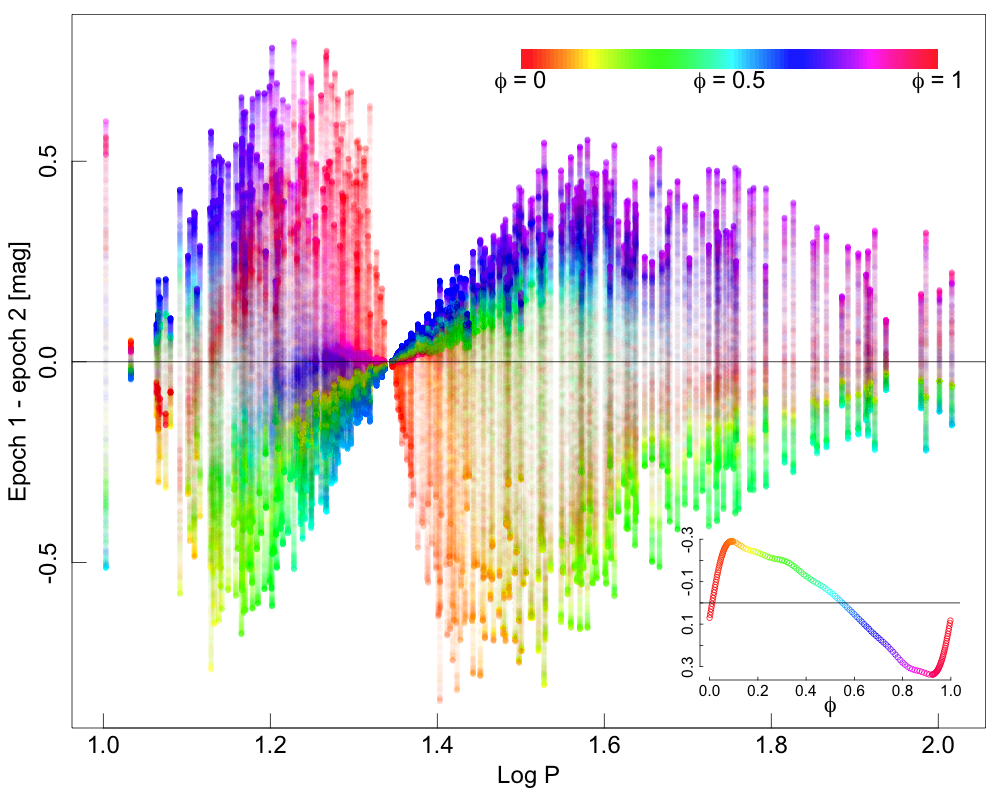}
\caption{\label{fg:phaseing} Graphical representation of the correspondence between Cepheid phase (color bar) and magnitude difference between epochs for F090W.  NGC$\,$4258 is on the left and NGC$\,$5584 on the right.}
\end{figure}

Comparing the two measurements in each filter with the predicted light curve for each Cepheid allows us to constrain to some extent their phases at the epoch of observations, and thus the phase correction, i.e., the difference between the measured magnitudes and the mean magnitude of each Cepheid.  The phase behavior of the epoch differences versus period are shown in \ref{fg:phaseing}.  We use the $I$-band light curve templates from \cite{Yoachim:2009} for  $F090W$ and the $H$-band template from \cite{Inno:2015} for the $F150W$ epochs, each with a small amplitude rescaling (5\% down from the $I$-band for F090W and 5\% up for $F150W$ from the $H$-band, estimated from the empirical relation between Cepheid amplitude and wavelength) appropriate for the difference in bandpass.  The amplitudes and periods of each Cepheid are based on their {\it HST} observations (R22); we estimate the amplitude in the $I$-band to be 0.6 times that in the visual, and use Eq.~12 of \cite{Riess:2020} to estimate the amplitude in the $H$-band.  We assume that the phase difference at maximum light scales as the logarithm of the period, with the $ H $-band maximum occurring $0.25+0.1(\log P-1)$ later than the $ I$-band maximum; this expression is based on fits of the light curves of Milky Way Cepheids. With this expression, the mean phase difference of our Cepheids is $\sim$0.3, consistent with the findings of \cite{soszynski05}.  We found similar phase corrections using the formulation of the phase difference from \citep{Inno:2015}.

\clearpage

\clearpage

To determine the probability distribution of the phase for each Cepheid, we start from a flat prior in phase, since any phase information from the $ HST $ observations is lost.  For each possible phase, we compare the magnitude difference between the two epochs with the difference predicted by the scaled templates, and assign the phase a relative probability based on how well the observed difference matches its predicted value, taking into account both measurement errors and intrinsic uncertainty in the templates.  This probability distribution, computed jointly for both filters, is then used to weight the inferred mean magnitude for that Cepheid.  The uncertainty in the mean magnitude includes both photometry errors and the probability distribution in phase.
With little empirical knowledge of $F277W$ light curves beyond ground-based $K$-band, we assumed a simple scaling of 70\% of the $F150W$ amplitude.

\begin{figure}[t]
\includegraphics[width=\textwidth]{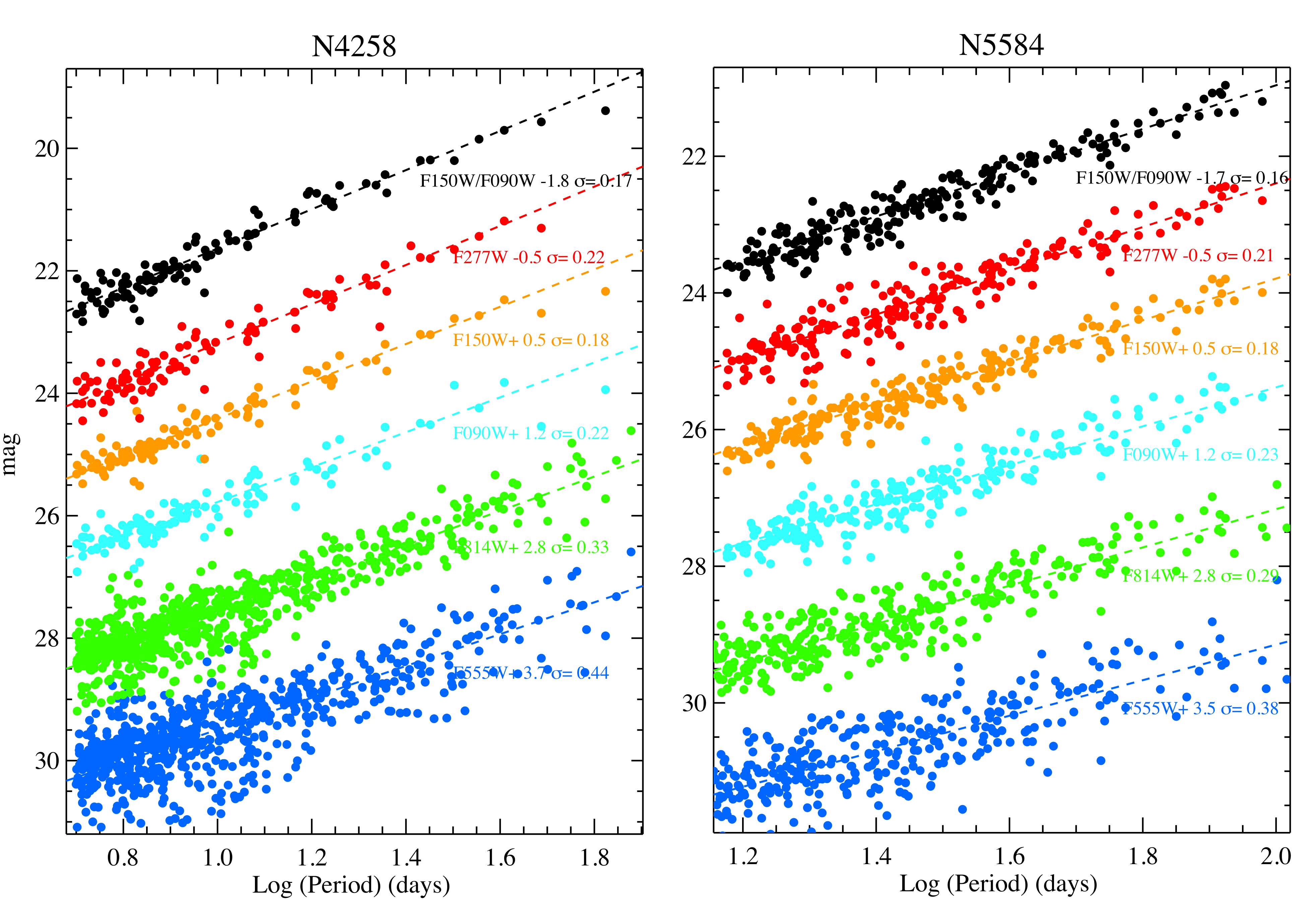}
\caption{\label{fg:pls} Period-luminosity relations from bluest (bottom) to least reddened (top).  The bottom two, $F555W$ and $F814W$ are from {\it HST} and the others from {\it JWST}.  The top set in black is a dereddened or Wesenheit magnitude, $F150W-R(F090W-F150W)$.  Magnitude offsets are applied as indicated for ease of view and the dispersion for each is given.}
\end{figure}

Fig.~\ref{fg:pls} presents single-band \PL relations from blue to red bandpasses.  These are improved upon in the next section with the simultaneous use of multiple bands to account for reddening.

We calculate an error estimate for each Cepheid magnitude as the quadrature sum of the photon statistics (as reported by DOLPHOT), the background error or crowding noise (as provided by the dispersion of the artificial stars around their mean), the width of the instability strip (see Figure \ref{fg:scatter}), and the phase uncertainty which varies from a random phase error of 0.14 mag~for $F150W$ (for Cepheids with only a single epoch or a period near a multiple of the epoch separation) to $\sim$0.03~mag for a well-constrained phase. The combined error varies by a factor of $\sim$3, from $\sim$0.1 to $\sim$0.3~mag in $F150W$ and a mean of 0.17-0.18~mag with the strongest determinants for the errors being the Cepheid location (which determines the crowding noise), period (which influences the crowding noise and the phase correction) and light curve sampling. The {\it JWST} crowding noise is a factor of $\sim$3.5 times smaller than for {\it HST} at the same NIR wavelength and is a minority of the combined noise term.  The errors are provided in Table \ref{tb:phot}.  

\subsection{Reddening-corrected Period-magnitude Relation\label{sec:wes}}

It is common to define ``Wesenheit'' \citep{madore82} indices to compare Cepheid magnitudes,
\begin{equation}
m^W_X=m_X-R(m_Y-m_Z)
\end{equation}
where $m_X$ is a magnitude in passband $X$, $m_Y-m_Z$ is a color (difference in two bandpasses), and $R$ is chosen to be the ratio of extinction in band $X$ for a change in color $Y-Z$ calculated from an extinction law.  Wesenheit indices serve to minimize the impact of extinction and finite width in temperature of the instability strip, reducing the intrinsic dispersion and providing the means to measure relative distances as long as a consistent value of $R$ is used.  We derive $R$ from the \cite{Fitzpatrick:1999} reddening law (with $R_V=3.3$ as used in R22).  

The Wesenheit index used for the last decade by the SH0ES team (e.g., R22) was $m_H^W=F160W- 0.39(F555W-F814W)$ based on these three {\it HST} bandpasses. The system we use for our {\it baseline} result here which we will refer to as the {\it JWST+HST} NIR system, is extremely similar to SH0ES {\it HST} system and is:
\begin{equation}
 (m_H^W)=F150W- 0.41(F555W-F814W). 
\end{equation}

The key difference is the substitution of {\it HST} WFC3-IR $F160W$ (mean wavelength 1.53$\mu$m) with {\it JWST} NIRCAM $F150W$ (mean wavelength 1.50$\mu$m; see Fig.~\ref{fg:filters}) and thus provides the most direct comparison to past {\it HST} measurements without the {\it HST} NIR confusion noise.  It uses the same {\it HST} color, $F555W-F814W$, which is well-measured (thanks to the high resolution of {\it HST} WFC3-UVIS and the high contrast between Cepheids and red giants in the optical) and provides high leverage on reddening (i.e., the lowest value of $R$ in the NIR).  In \S3.6 we will consider other bandpass combinations to form Wesenheit magnitudes.

It is important to measure the intercepts of the period-magnitude relations between hosts using the {\it same} \PL slope to equate their difference as their relative distance (although it is also possible to bin Cepheids in period and compare magnitudes at similar period, accepting the loss of information via binning).  We expect the $F150W$-based slope to be a bit shallower than $F160W$ (with slopes of -3.26 and -3.30 found by R16 and R22) due to the shorter effective wavelength and larger value of $R$.  Here we use the mean of our two hosts, -3.2.  We expect the empirical knowledge of the slopes in this bandpass to improve with future {\it JWST} observations of additional Cepheid hosts\footnote{At present the JWST Cepheid sample is too small to robustly compare individual host \PL slopes as the inclusion or exclusion of a single Cepheid, in certain cases, can produce larger differences than the statistical uncertainty of the included sample.  The best example is the longest period Cepheid in NGC$\,$4258 at log P $\sim$1.8 which is fainter than the \PL, in good agreement with the HST measurement of this same Cepheid, but the HST sample is larger because it has more fields and so it includes two at even longer periods that sit above the line, balancing around the nominal slope.  The greater uncertainty is shown by bootstrap resampling tests (R22) with uncertainties larger by 35\% for \PLs for cases of infrequent sampling at large periods such as for the LMC. Including or excluding this one point would change the JWST slope by $\sim$0.07 and toggle between the same slope as NGC$\,$5584 or a shallower slope, though its inclusion or exclusion has little impact on the relative intercepts.  The greater HST sampling at these periods shows the slopes to be equivalent (R22, see Fig.~10).}.  We also consider fits with a steeper and shallower slope by $\pm$0.05~mag/dex.

 \subsection{Baseline Results}  

We determine the intercepts within the {\it JWST+HST} NIR magnitude-system \PLs from the weighted mean after applying an iterative 3$\sigma$ clip. Accounting for a different level of uncertainty for individual measurements, this translates into a $\chi^2$ limit of 9 and excludes $\sim$2\% of the Cepheids. We also provide results with no Cepheid clipping.  

We derived a distance of NGC$\,$5584 from the two step ladder by adding the geometric distance modulus for NGC$\,$4258, $\mu_{0,N4258}=29.398$~mag \citep{Reid:2019} to the intercept difference between NGC$\,$5584 and NGC$\,$4258, i.e.,
\begin{equation}
\mu_{0,N5584}=m_H^{W,N5584}-m_H^{W,N4258}+\mu_{0,4258}.
\end{equation}
where $m_H^{W,N5584}$ and $m_H^{W,N4258}$ are the intercepts of their hosts linear \PL relations.

Our baseline distance modulus to NGC$\,$5584 is $\mu_0=31.813\pm0.020$~mag (not including the geometric distance uncertainty for NGC$\,$4258 of $\pm 0.032$). We can compare this to the {\it HST} NIR (SH0ES) result of $\mu_0=31.792 \pm 0.038$~mag also using the same geometric calibration from NGC$\,$4258.  (We note that SH0ES system distances are greater by $\sim$0.01~mag when calibrated by NGC 4258 as opposed to the mean of all three geometric anchors in R22).  The difference in distance is 0.01$\pm$0.04~mag, showing very good agreement.  Although the {\it HST} sample includes more Cepheids (560 vs. 325) due to the greater number of fields in NGC$\,$4258, the lower noise of the {\it JWST} data yields an error of about half the size as that of {\it HST}. In Table~\ref{tb:fits} we provide results of these \PL fits and parameters, including their dispersions.

The application of phase corrections (versus simple epoch averaging) reduces the \PL dispersion from 0.195 to 0.17~mag in both hosts. These gains are quite valuable, equivalent to a 30\% increase in sample size in terms of the error in the sample mean. As shown in Fig.~\ref{fg:scatter}, further gains in dispersion are possible with better constraints on the phases which would require additional epochs of {\it JWST} or {\it HST} observations obtained within $\sim$one year of these observations.  We have performed simulations that show a phase uncertainty of $\leq$0.1 would still significantly reduce the remaining phase noise and the resulting \PL dispersion to $\sim$0.13 mag, equivalent to an additional 70\% gain in sample size.  

The measured {\it distance} to the SN Ia host depends on {\it difference} in Cepheid magnitudes between hosts as shown in equation 3 and therefore does not depend on the Wesenheit (or telescope) system employed.  However, to directly compare the \PL {\it intercepts} (to R22 or other sources) which depend on absolute magnitudes, it is necessary to consider several transformations to account for subtle differences in the bandpasses and in the metallicities of the samples.  For both the {\it HST} and {\it JWST} samples we apply the same metallicity-dependence term by adding 0.21$\times$[O/H] as given in R22 (where [O/H] is the difference in metallicity from the solar value).  The mean metallicity for the Cepheids in both NGC$\,$4258 and NGC$\,$5584 as measured from their HII gradients (R22) happens to be the same at $[O/H]\sim$-0.10~dex, so the metallicity correction (from solar) is 0.02~mag in either case.  Since the mean metallicity for NGC$\,$4258 and NGC$\,$5584 is the same, this metallicity correction has no effect on the distance measure to NGC$\,$5584.  Rather, we apply it to ensure uniformity when comparing to the {\it HST} intercepts from R22 which sample a broader range of metallicities (such as the outer region of NGC$\,$4258).  We also include results without the metallicity correction.

We apply a color correction to the {\it HST} NIR magnitudes to account for the small difference between the $F150W$ and $F160W$ bandpasses (see Fig.~\ref{fg:filters} which we derive from the Padova stellar isochrones \citep{Bressan:2012} in the color range of Cepheids to be $F150W-F160W=0.033+0.036(F555W-F814W-1)$.  In future work we will use NIR spectrophotometry of MW Cepheids from {\it HST} to derive a fully-empirical measure of this transformation. As most Cepheids have $F555W-F814W$ close to 1, this term has a fairly constant value of $\sim$0.035~mag to within a few millimags.
   
We correct the {\it HST} intercepts for count-rate non-linearity (CRNL) by subtracting 0.035~mag for each following \cite{Riess:2019b}.  We note that we do not (yet) know the level of CRNL affecting NIRCAM, except to note that it is expected to be less than {\it HST} WFC3-IR (a H1RG device operated at temperatures above 110 K) because it uses a later detector generation \citep[H2RG run at 40 K,][]{Biesiadzinski:2011}.  Detector testing has shown these devices to have an order of magnitude fewer traps than WFC3-IR with trapping showing a direct correspondence to CRNL (Mike Regan, private communication). For this reason we will assume the {\it JWST} CRNL is much smaller than for {\it HST} WFC3-IR, i.e., $\leq 0.01$ mag, and make no adjustment to the {\it JWST} photometry until on-orbit measurements are completed.  Both the CRNL term and the color transformation cancel when we compare the intercepts of NGC$\,$5584 and NGC$\,$4258 and thus have no affect on the distance determination for NGC$\,$5584.
   
\begin{figure}[t]
\begin{center}
\includegraphics[width=0.85\textwidth]{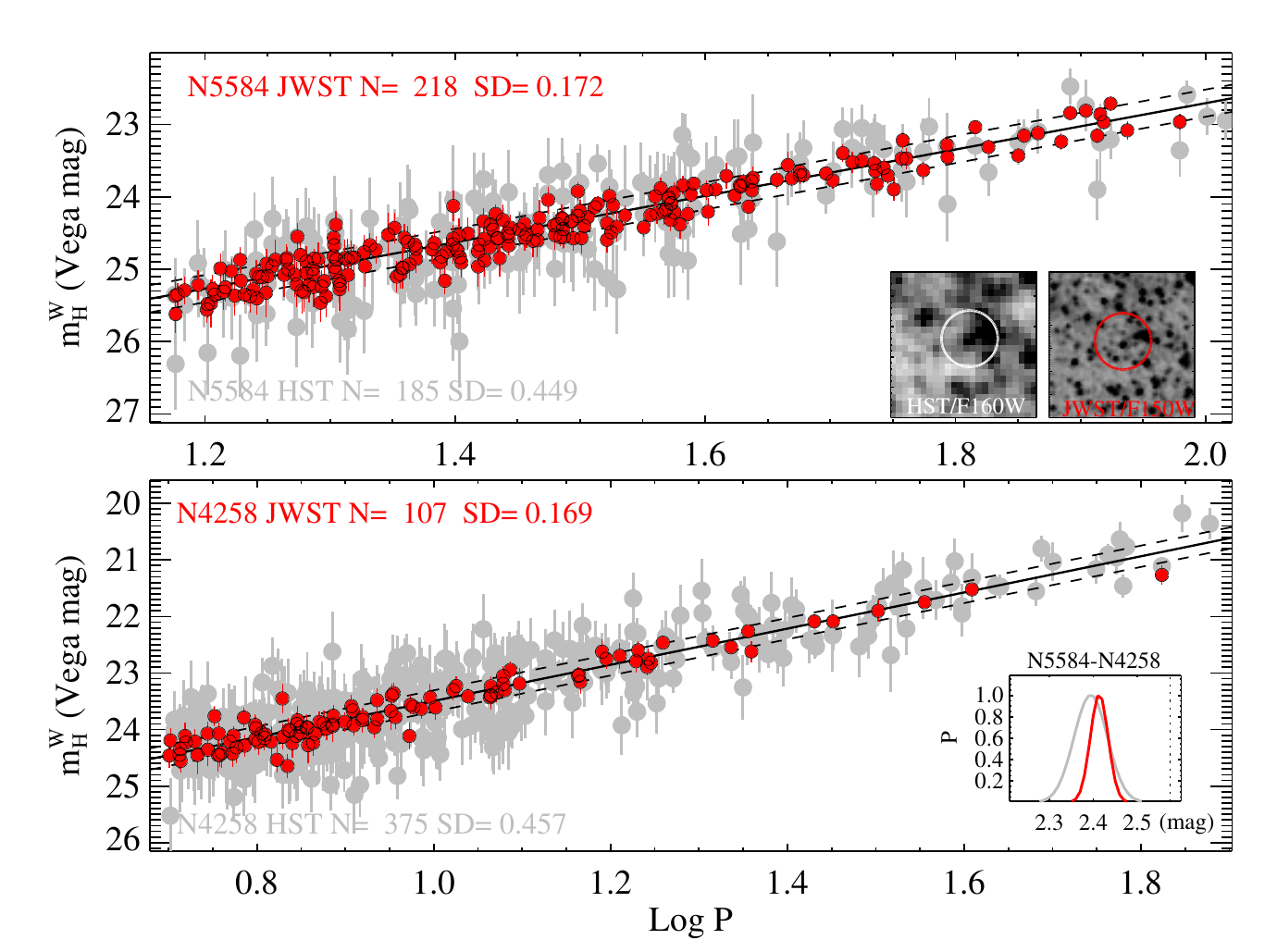}
\end{center}
\caption{\label{fg:comphstjwst} Comparison between the standard (SH0ES: R22) magnitude $W^H_{V,I}$ period-magnitude relation used to measure distances.  The red points use {\it JWST} $F150W$ and the gray points are from {\it HST} $F160W$, including a small transformation $F150W-F160W=0.033+0.036[(V-I)-1.0]$. The upper panel is NGC$\,$5584 with inset showing image stamps of the same Cepheid seen in the $H$-band by each telescope. The lower panel is NGC$\,$4258, with the inset showing the difference in distance moduli between NGC$\,$5584 and NGC$\,$4258 measured with each telescope.}
\end{figure}

The baseline intercepts from {\it JWST+HST} NIR are in good agreement with their color-transformed {\it HST}-based counterparts, with a difference ({\it JWST-HST}) of 0.02$\pm$0.03~mag for NGC$\,$5584 and 0.00$\pm$0.03~mag for NGC$\,$4258 as shown in Fig.~\ref{fg:comphstjwst}. Any CRNL or temporal degradation of {\it JWST} would go in the direction of reducing this mean difference in the intercepts of the absolute photometry (see \S 4).  

\subsection{Baseline Variants}

In Table \ref{tb:fits} we present alternatives to the baseline fit with the nature of the variation given in column 2.  These include no phase correction (i.e., a simple average of epochs), applying chip-to-chip offsets derived from non-variable stars in overlapping chips, limiting all Cepheids in both hosts to $P> 15$ days, no Cepheid outlier rejection, steeper and shallower \PL slopes, limiting Cepheids to those present in the other telescope sample, and forgoing the metallicity correction. Changing the slope by $\pm 0.05$ changes the distance to the SN Ia host by $\pm 0.024$ mag, but a similar is seen for the {\it HST} sample so the agreement in distance remains $<0.015$ mag.  Chip-to-chip offsets have little impact, raising the distance by 0.012~mag if applied, a consequence of the wide distribution of Cepheids across NIRCAM chips and modules. Limiting both host samples to $P>15$ days (which excludes Cepheids in NGC$\,$4258 with $P<15$ days) produces no change in distance but raises the uncertainty by 50\%.  Limiting the JWST sample to Cepheids observed by {\it HST} F160W increases the distance by $\sim$0.01~mag and increases the uncertainty by 25\%.  We also split the sample into equal weight halves divided according to the level of background seen in the {\it JWST} images (based on the mean of the artificial stars, not the sources to avoid a selection bias).  The division is unequal in terms of the number of Cepheids, two-to-one for high-to-low background, because the Cepheids with low background have greater weight.  The difference in the distance to NGC 5584 for each half-weight sample is not significant (0.5 $\sigma$). The sense of the difference would be of lower crowding backgrounds reducing the distance and raising the $H_0$ tension, but again the difference is not significant.
    
For the {\it HST} sample we calculate similar variants useful for comparisons.  We also split the {\it HST} sample into equal ({\it HST}) weight halves divided according to the level of crowding seen in the {\it JWST} images (calculated from the {\it JWST} artificial stars). 
The difference in the distance to NGC 5584 between the two {\it HST} sample halves is not significant (0.2$\sigma$) with the precision of this particular comparison limited by the higher dispersion of the {\it HST} \PLs and the small spatial overlap of the {\it JWST} fields with the six {\it HST} fields from which Cepheids were mined from NGC$\,$4258 \citep{Yuan:2022a}.  The sense of the difference would be of lower crowding backgrounds reducing the distance and raising the $H_0$ tension, but again the difference is not significant and much less powerful for evaluating the {\it HST} photometry than by comparing the full samples from each telescope. This is consistent with the finding in R22 (see Figure B2 there).   We discuss issues related to selecting Cepheid subsamples and potential bias or loss of information in \S 4.1.

As all variants and tests yield results consistent with the baseline, we consider the baseline results robust.

\subsection{Different Filter Combinations}

We calculate different Wesenheit indices using Cepheid photometry in three {\it HST} WFC3 bands, $F555W$, $F814W$ and $F160W$ (R22) and three {\it JWST} NIRCAM bands, $F090W$, $F150W$, and $F277W$ (see Fig.~\ref{fg:filters}). We will consider five period-magnitude Wesenheit relations which appear most informative of the {\it JWST} data using the following definitions:

\begin{itemize}
    \item {\it JWST+HST} NIR $(m_H^W): F150W- 0.41(F555W-F814W)$
    \item {\it JWST} NIR: $F150W - 0.72(F090W-F150W)$
    \item {\it JWST+HST} MIR: $F277W- 0.17(F555W-F814W)$
    \item {\it JWST} MIR: $F277W- 0.30(F090W-F150W)$
     \item {\it JWST+HST} OPT: $F090W-0.98(F555W-F814W)$
   \end{itemize}

We use a mean slope of -3.3 for MIR or $F277W$ systems found for these two hosts, similar as seen from ground for K-band systems \citep{persson04}.  
  
The resulting distance moduli for NGC$\,$5584 range from 31.806 to 31.839 mag, consistent with the baseline result.  More noteworthy are differences in their \PL dispersions.  Using only {\it JWST} NIR data yields similar dispersion as the baseline system.  Going to either redder wavelengths ($F277W$ in place of $F150W$) or bluer ($F090W$ in place of $F150W$) yields a larger dispersion of $\sim$0.20~mag vs. 0.17~mag as shown in Fig.~\ref{fg:scatter}, likely due to added crowding in the case of $F277W$ and greater extinction (residuals) for $F090W$.  Overall, the most precise distance indicators appear to be the {\it JWST} NIR and JWST+HST NIR systems.

The \PL including $F277W$ can be used, in principle, to investigate differences in individual Cepheid metallicity as this band overlaps with the wide molecular CO absorption in the Cepheid photosphere from 2.3 to 3.0$\mu$m \citep{Scowcroft:2016}.  It is useful to compare the {\it JWST} MIR \PL to {\it JWST} NIR as the primary difference is inclusion of $F277W$ in the former.  Their difference in distance moduli is 0.01$\pm$0.03~mag which is not significant.  This is not surprising since both galaxies have the same metallicity (according to HII measurements and indicated by the host masses).  However, if the Cepheid metallicities were extremely different, by a dex (e.g., the difference in very metal rich Cepheids in the local solar neighborhood and those in the SMC), a difference of $\sim$0.04~mag may be expected and which might be detectable with a larger sample of hosts.

\begin{figure}[b]
\begin{center}
\includegraphics[width=0.6\textwidth]{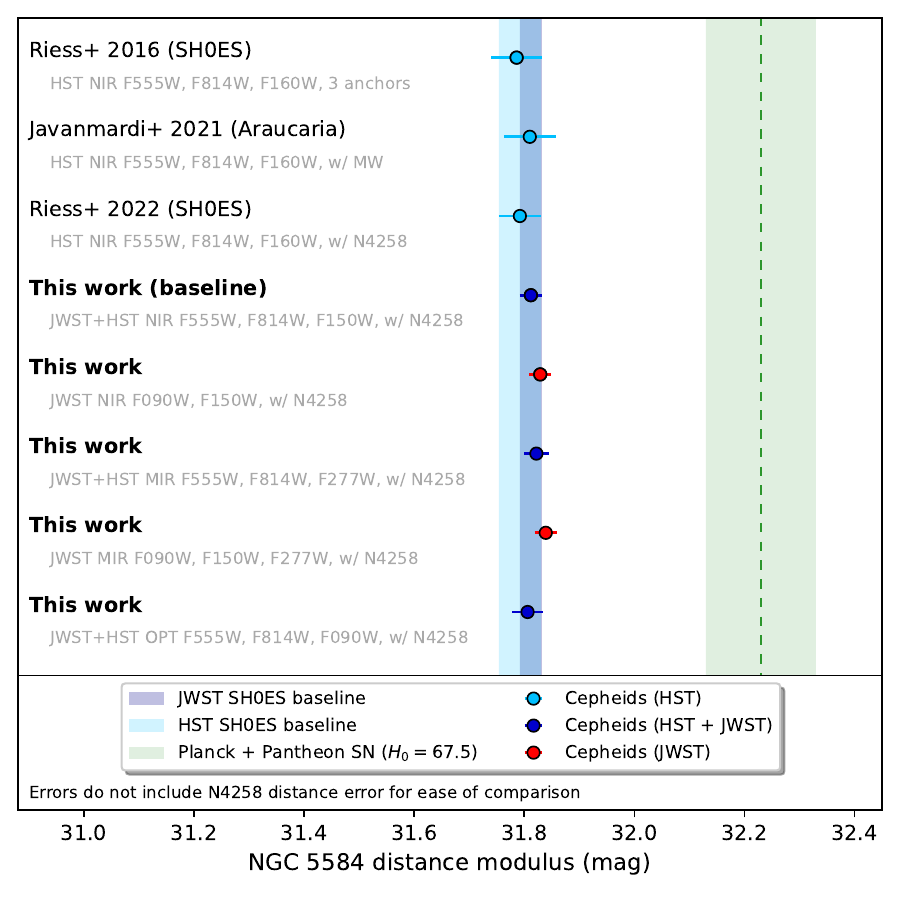}
\end{center}
\caption{\label{fg:distances} A comparison of Cepheid distance measures to NGC$\,$5584.}
\end{figure}

Fig.~\ref{fg:distances} compares the various results discussed in this section with previous Cepheid-based distances obtained from {\it HST-}only measurements.

\clearpage

\begin{deluxetable}{llr|rlrr|rlrr|lr}[t]
\tablenum{3}
\tablecaption{ {\it HST}  and {\it JWST} Cepheid \PL Fits  \label{tb:fits}}
\tablewidth{0pt}
\tablehead{\colhead{Sample} & \colhead{comment} & \colhead{PL} & \multicolumn{4}{c}{NGC$\,$5584} & \multicolumn{4}{c}{NGC$\,$4258} & \multicolumn{2}{c}{NGC$\,$5584} \\[-9pt]
\colhead{} & \colhead{} & \colhead{slope} & \multicolumn{1}{c}{{\scriptsize N}} & \multicolumn{1}{c}{{\scriptsize{zp}}} & \multicolumn{1}{c}{{\scriptsize $\sigma$}} & \multicolumn{1}{c}{{\scriptsize SD}} & \multicolumn{1}{c}{{\scriptsize{N}}} & \multicolumn{1}{c}{{\scriptsize ZP}} & \multicolumn{1}{c}{{\scriptsize $\sigma$}} & \multicolumn{1}{c}{{\scriptsize{SD}}} & \multicolumn{1}{c}{{\scriptsize $\mu$}}  & \multicolumn{1}{c}{{\scriptsize $\sigma^c$}}}
\startdata
\tableline
\multicolumn{13}{c}{{\bf Baselines}} \\
\tableline
HST NIR (SH0ES) & $F555W,F814W,F160W$  & -3.20  &  185  &  29.089 $^b$ &  0.032  &  {\bf 0.449}  &  375  &  26.694 $^b$ &  0.020  &  {\bf 0.457}  &  {\bf 31.792} $^a$ &  {\bf 0.038}  \\ 
JWST+HST NIR &   $F555W,F814W,F150W$ &  -3.20 &  218  &  29.107  &  0.010  &  {\bf 0.172}  &  107  &  26.694  &  0.017  &  {\bf 0.169}  &  {\bf 31.813}  &  {\bf 0.020}  \\ 
\tableline
\multicolumn{13}{c}{Baseline Variants} \\
\tableline
JWST+HST NIR & no phase correction  & -3.20  &  221  &  29.119  &  0.014  &  0.195  &  109  &  26.698  &  0.021  &  0.194  &  31.819  &  0.025  \\ 
JWST+HST NIR &  chip-to-chip offsets  & -3.20  &  217  &  29.112  &  0.010  &  0.171  &  107  &  26.686  &  0.017  &  0.170  &  31.824  &  0.020  \\ 
JWST+HST NIR &  $P>15$ days & -3.20  &  218  &  29.107  &  0.010  &  0.172  &   19  &  26.699  &  0.030  &  0.137  &  31.807  &  0.032  \\
JWST+HST NIR &  no $\sigma$-clip & -3.20  &  223  &  29.107  &  0.010  &  0.211  &  113  &  26.678  &  0.017  &  0.227  &  31.827  &  0.020  \\ 
JWST+HST NIR &  steeper slope & -3.25  &  217  &  29.188  &  0.010  &  0.170  &  106  &  26.752  &  0.017  &  0.163  &  31.834  &  0.020  \\ 
JWST+HST NIR &  shallower slope & -3.15  &  218  &  29.029  &  0.010  &  0.172  &  107  &  26.641  &  0.017  &  0.167  &  31.787  &  0.020  \\ 
JWST+HST NIR &  steeper and $P>15$ days & -3.25  &  217  &  29.188  &  0.010  &  0.170  &   19  &  26.767  &  0.030  &  0.140  &  31.818  &  0.032  \\
JWST+HST NIR &  and in SH0ES $F160W$  & -3.20  &  157  &  29.087  &  0.012  &  0.166  &   62  &  26.665  &  0.023  &  0.177  &  31.821  &  0.026  \\ 
JWST+HST NIR &  no metallicity cor.  & -3.20  &  218  &  29.133  &  0.010  &  0.170  &  107  &  26.711  &  0.017  &  0.169  &  31.819  &  0.020  \\ 
JWST+HST NIR & high {\it JWST} background & -3.20  &  135  &  29.122  &  0.016  &  0.184  &   72  &  26.694  &  0.024  &  0.169  &  31.826  &  0.029  \\
JWST+HST NIR & low {\it JWST}  background & -3.20  &   82  &  29.101  &  0.014  &  0.158  &   35  &  26.693  &  0.025  &  0.168  &  31.806  &  0.029  \\ 
\tableline
HST NIR & steeper slope & -3.25  &  185  &  29.168 $^b$ &  0.032  &  0.449  &  375  &  26.751 $^b$ &  0.020  &  0.457  &  31.814 $^a$ &  0.038  \\ 
HST NIR  & shallower slope & -3.15  &  185  &  29.010 $^b$ &  0.032  &  0.450  &  375  &  26.638 $^b$ &  0.020  &  0.457  &  31.770 $^a$ &  0.038  \\ 
HST NIR  & and in {\it JWST} $F150W$ & -3.20  &  160  &  29.114 $^b$ &  0.035  &  0.451  &   65  &  26.744 $^b$ &  0.056  &  0.384  &  31.768 $^a$ &  0.066  \\
HST NIR  & high {\it JWST} background &  -3.20 & 117  &  29.159 $^b$ &  0.049  &  0.454  &   47  &  26.781 $^b$ &  0.083  &  0.403  &  31.776 $^a$ &  0.096  \\ 
HST NIR  & low {\it JWST} background & -3.20 &   43  &  29.065 $^b$ &  0.051  &  0.448  &   18  &  26.716 $^b$ &  0.076  &  0.323  &  31.747 $^a$ &  0.092  \\ 
HST NIR & no metallicity cor. & -3.20  &  185  &  29.114 $^b$ &  0.032  &  0.447  &  375  &  26.722 $^b$ &  0.020  &  0.457  &  31.790 $^a$ &  0.038  \\
\tableline
\multicolumn{13}{c}{Different Filter Combinations} \\
\tableline
JWST  NIR &  $F090W,F150W$ & -3.20  &  213  &  29.051  &  0.011  &  0.161  &  106  &  26.620  &  0.017  &  0.175  &  31.829  &  0.020  \\ 
JWST  MIR &  $F090W,F150W,F277W$  & -3.25  &  204  &  29.172  &  0.011  &  0.191  &   98  &  26.730  &  0.017  &  0.203  &  31.839  &  0.020  \\ 
JWST+HST MIR & $F555W,F814W,F277W$ & -3.25  &  209  &  29.187  &  0.012  &  0.201  &   98  &  26.763  &  0.018  &  0.194  &  31.822  &  0.022  \\ 
JWST+HST OPT & $F555W,F814W,F090W$ & -3.15  &  219  &  29.190  &  0.013  &  0.203  &  110  &  26.782  &  0.024  &  0.212  &  31.806  &  0.028  \\ 
\tableline
\enddata
\tablecomments{$^a$ R22 Table 6 $\mu_{N5584}=31.772 \pm 0.052$ based on three anchors and $P>18$ days, here we use only one anchor, NGC$\,$4258, and $P > 15$ days to allow a direct comparison. $^b$ To allow a direct comparison with JWST$+$HST NIR, we 
applied a transformation of $F150W-F160W=0.033+0.036(F555W-F814W-1)$ which adds 0.04 and 0.03 to NGC$\,$5584 and NGC$\,$4258, respectively, and corrected here for CRNL by the subtraction of 0.035~mag for both NGC$\,$5584 and NGC$\,$4258, the two corrections canceling to $< 0.01$~mag with no impact on distance. $^c$ error does not include geometric distance uncertainty for NGC$\,$4258 of $\pm 0.032$.}
\vspace{-36pt}
\end{deluxetable}

\section{Discussion}

The two galaxies studied here suffice to complete a distance ladder from geometry to one SN~Ia host.  But the more direct value of this study is in comparing the relative distances of these two galaxies obtained via Cepheid observations with $ JWST $ vs. $ HST $.  This comparison involves large Cepheid samples (325 with {\it JWST} and 560 with {\it HST}); because both are compared {\it directly} Cepheids-to-Cepheids and {\it internally} to their respective photometric systems, uncertainties related to Cepheid calibrations, photometric zero points, and SN~Ia properties are irrelevant in this comparison.  The number of Cepheids is sufficient to compare their relative distances to an accuracy $\sigma=0.04$ mag; our baseline analysis yields $ \mu_{N5584}-\mu_{N4258} = 2.413 \pm 0.020 $ with {\it JWST}, vs.~$2.395 \pm 0.038 $ with {\it HST}.  Since {\it JWST} yields IR photometry that is vastly less affected by crowding than {\it HST}, this provides the strongest indication to date that crowding does not play a role in the $\sim$0.18~mag Hubble Tension.  We also note that a full reanalysis of the {\it HST} photometry of the Cepheids in NGC$\,$5584 starting from the raw data and using different photometry tools was undertaken by \cite{Javanmardi:2021} who found $\mu=31.810 \pm 0.047$~mag (also correcting for crowding using artificial stars) in good agreement with the result from the SH0ES team (R22).  Thus the confirmation of {\it HST} Cepheid photometry by {\it JWST} is not particularly method-dependent. We anticipate ten more hosts of 15 more SN~Ia from among the SH0ES (R22) sample will be observed with {\it JWST} in these same filters in the next year in programs 1685 and 2875.

On a closer look, it is evident that if a significant error in {\it HST} Cepheid photometry associated with confusion existed, NGC$\,$5584 would have offered the best chance to discover it, for two reasons.  The first is that it has one of the highest combinations of distance and Cepheid sample size in R22 (the most distant host with $\geq$ 200 Cepheids) which maximizes the {\it leverage} for detecting a crowding-based offset. Its distance judged by the SN or the Cepheids places it within a few tenths of a magnitude of the SH0ES sample mean.  

In addition, the Cepheid distance to NGC$\,$5584 and the SN~Ia it hosted are $\sim$2-2.5 $\sigma$ ($\sim$0.25 mag) from the mean of the SH0ES sample of 42 SN~Ia vs Cepheid comparisons {\it in the direction where its Cepheids are bright for its SN~Ia standardized magnitude or the SN~Ia is faint for the host Cepheids}, i.e., in the direction of greater Hubble constant tension. Thus if a bias in {\it HST} photometry existed making Cepheids in most or any hosts appear too bright at a level comparable to the Hubble tension (i.e., $\geq$0.15~mag level), it is most likely present for the hosts well off the mean in this direction and detectable with a large Cepheid sample, two features which uniquely occur for NGC$\,$5584.  However, the agreement between the {\it JWST} and {\it HST} photometry indicates this is not the case.  A similar result was seen by \cite{Yuan2022JWST} for {\it JWST} observations of NGC$\,$1365 obtained for a different purpose, but the precision and leverage of the data here (including an anchor in NGC$\,$4258), makes this test far stronger.

Considering the SN~Ia, SN 2007af is spectroscopically and photometrically normal with low-reddening ($c\sim$ 0, $x1\sim$-0.4), and very well-observed with 4 independent light curve sets in the Pantheon$+$ compilation from the most prolific surveys (CSP, Cfa, SWIFT, KAIT) which all yield consistent results.  Therefore we have no reason to doubt the SN~Ia measurement or standardization. If this SN~Ia, calibrated by its host Cepheids, is $\sim$0.25~mag faint (after standardization) it would not represent a particularly surprising statistical fluctuation of SN~Ia magnitudes.  Standardized SN~Ia in the Pantheon$+$ sample \citep{Brout:2022} exhibit a dispersion of $\sim$0.13~mag in the Hubble flow and thus among 42 SN~Ia calibrated by Cepheids in the SH0ES sample, we expect a couple will lie as far as SN 2007af off the sample mean.  It is also similarly faint for its Cepheid distance in the NIR \citep{Galbany:2022}, so the faintness would be due to an intrinsic fluctuation in luminosity rather than due to dust.  

\subsection{Considerations for Cepheid Subsamples and Artificial Stars}

The {\it JWST} photometry presented here demonstrates the fundamental validity of the artificial star tests as used to correct for crowding in Cepheid photometry \citep{Riess:2012,Riess:2016,Riess:2022}.  A critical element of this approach, as discussed in \S 2.3, is that {\it any} selection criteria applied to a Cepheid population corrected statistically via artificial star tests {\it must} also be applied to the artificial stars themselves, or biases will appear, because the mean background indicated by the artificial stars will not match that which applies to the Cepheids. This issue can arise when selecting a subsample of Cepheids from a computed catalog of photometry. \footnote{An example of this is seen in \cite{Mortsell:2022}, who excluded the reddest Cepheid photometry in R16 but did not likewise exclude equally red artificial stars.  Because the background of NIR images are dominated by red giants, a red Cepheid color correlates to a high background;  their process results in an overestimate of the background subtraction, and thus a significant faintward bias in the derived Cepheid magnitudes.}

Whenever new selection criteria are considered, it is necessary to remeasure the photometry and characterize artificial stars applying such criteria consistently.  This exercise can only be done from the image pixels, and not from a photometry catalog such as R22, which already includes the application of artificial stars to account for crowding.   

Artificial star distributions serve a second function besides removing bias: they are used to estimate individual Cepheid uncertainties due to crowding noise and provide the relative weights of Cepheid photometry for fitting.  The weights, which vary due to the local scale of background fluctuations, make it unnecessary to select a golden sample of less crowded Cepheids, as more crowded Cepheids are appropriately down-weighted.  For example, it may appear helpful to use the superior resolution of the {\it JWST} images to identify and exclude Cepheids with companions that are not resolved at {\it HST} NIR resolution to reduce crowding noise in {\it HST} Cepheid photometry.  To follow this method without incurring a bias it would still be necessary to eliminate such contaminated sources consistently for both real and artificial stars, as explained in the prior paragraph; this would involve adding and scrutinizing artificial stars added in the same location to both {\it HST} and {\it JWST} images.  However, this approach would be equivalent to discarding data with a low signal-to-noise when calculating a weighted mean.  There is no formal advantage in discarding the high crowding sources. While they have less weight, they still contribute to increasing the signal-to-noise of the mean intercept and the apparent \PL dispersion will be properly quantified if similarly weighted.  This is apparent in our analysis when we split the {\it JWST}+{\it HST} NIR sample by their background levels as measured from artificial stars as shown in Table \ref{tb:fits}.  The low-background sample has a smaller \PL dispersion (0.158 vs 0.184~mag for NGC 5584) and equal weight with half as many Cepheids as the high background sample, but both together give the lowest intercept error of all (0.010~mag for NGC 5584). The same is seen for the {\it HST} sample split by {\it JWST} background; although the sample is much smaller (because it requires {\it JWST} imaging of the {\it HST} regions, the weight of the less crowded half comes from a third of the sample size.  Of course, this consideration requires the crowding correction is statistically unbiased, as proven by the {\it JWST } photometry.

In practice, crowding, like weight, is a continuum and less crowding provides higher weight.
The recovered artificial star dispersions would seem to offer the best guidance for the relative weights of all {\it JWST} measured Cepheids (``crowded'' or not) and avoids an arbitrary selection of a ``crowding limit''.   Further, the use of {\it JWST} as a veto for {\it HST} requires {\it JWST} images; as we have found here, the {\it JWST} photometry is more precise than {\it HST}, so, when possible, its direct use offers the greater value.  

\subsection{Outlook}

Other distance indicators such as the tip of the red giant branch (TRGB, \citealt{Lee1993}) and J-AGB stars \citep{Madore2020} are readily apparent in the same images of NGC$\,$5584 and NGC$\,$4258 used to measure the Cepheids.  We defer their analysis until we have the opportunity to better characterize these population-based indicators whose characterization is best served by a larger sample of hosts.  When the {\it JWST} sample of SN~Ia hosts grows to a significant fraction of the {\it HST} sample, both from this program and others underway (i.e., 1995 and 2875), we anticipate further improvements to the precision and accuracy of the measurement of the Hubble constant.

\begin{acknowledgments}
We are indebted to all of those who spent years and even decades bringing {\it JWST} to fruition. This research made use of the NASA Astrophysics Data System.  We thank Martha Boyer, Yukei Murakami and Siyang Li for helpful conversations related to this work.  We thank our PC Alison Vick.

G.S.A acknowledges support from JWST GO-1685.  RIA is funded by the SNSF via an Eccellenza Professorial Fellowship PCEFP2\_194638 and acknowledges support from the European Research Council (ERC) under the European Union's Horizon 2020 research and innovation programme (Grant Agreement No. 947660).

\end{acknowledgments}
\bibliographystyle{apj} %
\bibliography{ms}
\end{document}